\documentclass[
twocolumn,
floats,floatfix,
nofootinbib,
prd,aps,superscriptaddress,tightenlines]{revtex4}
\usepackage{graphicx}
\usepackage{bm}
\usepackage{pstricks}

\newcommand{\xslash}[1]{{\rlap{$#1$}/}}
\def\vev#1{\left\langle #1 \right\rangle}
\newcommand{\me}[3]{\ensuremath{\left\langle{#1}\vphantom{#2 #3}
\right|{#2}\left|\vphantom{#1 #2}{#3}\right\rangle}}

\def\OMIT#1{{}}
\def\lqcd{\Lambda_{\rm QCD}}

\def\dist{\mu}
\newcommand{\nn}{\nonumber \\ }
\newcommand{\beq}{\begin{equation}}
\newcommand{\eeq}{\end{equation}}
\newcommand{\beqa}{\begin{eqnarray}}
\newcommand{\eeqa}{\end{eqnarray}}

\newcommand{\bn}{{\bar n}}

\def\bN{\bar N}

\begin{document}

\title{Shape function effects in $B \to X_s \gamma$ and $B \to X_u \ell \bar \nu$ decays}

\author{Christian W.~Bauer}
\affiliation{California Institute of Technology, Pasadena, CA 91125}

\author{Aneesh V.~Manohar}
\affiliation{Department of Physics, University of California at San Diego,
  La Jolla, CA 92093}
\date{December 2003}

\begin{abstract}
We calculate the decay distributions for inclusive $B \to X_s \gamma$ and $B \to X_u \ell \bar \nu$ decays in the endpoint region, where radiative corrections are large. The computation is done using effective field theory methods. The matching coefficients are computed to ${\cal O}(\alpha_s)$, and the anomalous dimensions to next-to-leading order. The final expressions for the differential decay spectra include the complete ${\cal O}(\alpha_s)$  corrections, and sum the leading and next-to-leading Sudakov series. We present results for regions of phase space where the shape function can be expanded in local operators, and give the matching coefficients of the resulting enhanced non-perturbative effects to order $\alpha_s$. We show that moments of the shape function {\it are not} given by moments of local operators once perturbative effects are included,  explain why the shape function and its moments satisfy different renormalization group equations, and contrast this with the situation for deep inelastic scattering. We  show that there are large perturbative corrections in the usual definition of the shape function. This renders incorrect previous prescriptions for combining radiative corrections with the shape function.
\end{abstract}

\maketitle

\section{Introduction}

The inclusive decay rates and differential decay distributions for $B \to X_s \gamma$, $B \to X_s e^+ e^-$, and $B \to X_u \ell \bar \nu$  can be computed~\cite{inclusive,MW,bsgOPE,book} in a systematic expansion in powers of $\lqcd/m_b$ and $\alpha_s(m_b)$ using heavy quark effective theory (HQET)  and the operator product expansion. The operator product expansion for the differential decay distribution fails in certain kinematic regions, but one can still compute a suitably smeared decay distribution. In $B \to X_s \gamma$ decay, the operator product expansion gives an expansion for the photon spectrum ${\rm d}\Gamma/{\rm d}E_\gamma$ containing derivatives of $\delta$-functions of the form $\lqcd^n \delta^{(n)}(m_b - 2 E_\gamma)$, which are singular in the endpoint region $2E_\gamma \sim M_B$. Nevertheless, one can still compute the differential decay rate in the endpoint region provided one smears over a range of photon energies which is large compared with $\lqcd$. Smearing over a range of photon energies of order $\Delta$ converts the expansion of ${\rm d}\Gamma/{\rm d}E_\gamma$ from an expansion in singular terms to an expansion in powers of $(\lqcd/\Delta)^n$.\footnote{See, for example, Ref.~\cite{LLMW} where smeared moments for $B \to X_s \gamma$ are computed to order $\alpha_s^2 \beta_0$.} In the resonance region defined by $\Delta \sim \lqcd^2/M_B$, the operator product expansion breaks down. In this region the invariant mass of the final hadronic system satisfies $M^2_X \sim \lqcd^2$, and the inclusive $B \to X_s \gamma$ rate is computed by summing over form-factors for a few exclusive modes. 

In the shape function region defined by $\Delta\sim \lqcd$, all the $(\lqcd/\Delta)^n$ terms are equally important. The most singular terms $\lqcd^n \delta^{(n)}(m_b - 2 E_\gamma)$ can be summed into a non-perturbative shape function~\cite{shape}  $S(E_\gamma)$ that describes the photon spectrum in the endpoint region. The subleading singularities, $\lqcd^n \delta^{(n-1)}(m_b - 2 E_\gamma)$ give rise to subleading shape functions suppressed by $\lqcd/m_b$~\cite{subshape}, and will be neglected here. The final state invariant mass in the shape function region is $M^2_X \sim M_B \lqcd \ll M_B^2$, so the final state is jet-like. In this kinematic region, the final state is described by a collinear-quark, and the appropriate effective theory to use is soft-collinear effective theory (SCET)~\cite{BFL,BFPS,BS,BPS}. 

The shape function region is important for measurements of $B \to X_s \gamma$ and $B \to X_u \ell \bar \nu$ decays, because experimental cuts needed to eliminate backgrounds restrict the measurable region of the allowed phase space. The shape functions for $B \to X_s \gamma$ and $B \to X_u \ell \bar \nu$ decays
 are related, so the shape function can be measured in one process and then used in the other~\cite{Falk}. This avoids introducing model dependence into the analysis of experimental results. QCD radiative corrections are large in the shape function region, due to Sudakov double logarithms, which need to be summed. The effect of Sudakov resummation in $B$ decays, and its effects on the extraction of CKM parameters such as $V_{ub}$ has been studied in detail previously in a series of papers by Leibovich, Low and Rothstein~\cite{Leibovich}.
 
In this paper, we will study the shape function region in $B \to X_s \gamma$ and $B \to X_u \ell \bar \nu$ using SCET. Our results are given for arbitrary $q^2$, and so can  be used for $B \to X_s e^+ e^-$ decay. The weak decay Hamiltonian for $B \to X_s e^+ e^-$ is considerably more complicated than for the other two cases, and is a sum over the different operators. The expressions for $B \to X_s e^+ e^-$ can be obtained using the methods of this paper, and are not given explicitly.
 
Another region of interest is when the smearing size $\Delta$ is large compared with $\lqcd$, $m_B \gg \Delta \gg \lqcd$. In this case, one has an expansion in terms of local operators. Some of the non-perturbative corrections are enhanced, and are of order $(\lqcd/\Delta)^m$, rather than $(\lqcd/m_b)^m$. These enhanced non-perturbative corrections will also be computed in this paper.

The SCET renormalization group equations allow one to sum the Sudakov double logarithms in the endpoint region. The matching conditions are computed to order $\alpha_s$, and the renormalization group evolution to next-to-leading order. The matching conditions and anomalous dimensions for the first part of our calculation have been computed previously~\cite{BFL}. We extend the results to include all non-vanishing terms for large moments.  In the regime $\Delta \gg \lqcd$, the total order $\alpha_s$ part of the calculation (i.e.\ not including renormalization group evolution) agrees with an existing calculation by de~Fazio and Neubert~\cite{dFN}. The total order $\alpha_s$ contribution is generated at different scales. Our calculation shows that the most singular order $\alpha_s$ terms, of the form $\ln(1-x)/(1-x)$ are generated at two different scales, and some of them are included in the shape function.  As a result, shape function effects cannot simply be incorporated by convoluting a non-perturbative shape function with the order $\alpha_s$ decay distributions, as was done in Ref.~\cite{dFN}, and we disagree with their results in the shape function region $\Delta \sim \lqcd$.

We study the connection between the shape function and its moments. Unlike the case of deep inelastic scattering, the shape function and its moments satisfy different renormalization group evolution, and are not simply related. The differences arise because of the existence of the velocity $v$ in heavy quark decays, which couples the $+$ and $-$ lightcone components of momentum, and because the twist expansion of deep inelastic scattering is not valid for heavy hadron decays. This implies that moments of the shape function are not related to matrix elements of local operators as was previously assumed.

The outline of the paper is as follows. The kinematics and notation are summarized in Section~\ref{sec:kin}, and the important mass scales  in Section~\ref{sec:mass}. The decay distributions and  hadronic tensors that we compute are discussed in Section~\ref{sec:htensor}. The details of the computation are given in Section~\ref{sec:pert}, and the relation between the shape function and its moments in Section~\ref{sec:local}. Some applications are discussed in Section~\ref{sec:app}. The reader not interested in theoretical details can skip directly to section~\ref{sec:app}.

\section{Notation and Kinematics}\label{sec:kin}

The velocity of the decaying $B$ meson is $v^\mu$, and we will work in the rest frame $v^\mu=(1,0,0,0)$.
For $B \to X_s \gamma$, $B \to X_s e^+ e^-$, and $B \to X_u \ell \bar \nu$ decay, the $b$ quark  decays into a light quark with momentum $p$ and a gauge boson with momentum $q$. The coordinate axes are chosen so that the virtual gauge boson is emitted in the $z$-direction, and the light quark in the negative $z$-direction. It is convenient to define the null vectors $n^\mu=(1,0,0,-1)$ and $\bar n=(1,0,0,1)$, which satisfy
\begin{eqnarray}
2v^\mu = n^\mu + \bn^\mu\,, \qquad n^2 = \bn^2 = 0 \,, \quad n \cdot \bn =2.\nonumber
\end{eqnarray}
Any four-vector $a^\mu$ can be written as
\begin{eqnarray}
a &=& \frac 1 2 a^+ \bn^\mu + \frac 1 2 a^- n^\mu + \mathbf{a}_\perp,\nonumber
\end{eqnarray}
where
\begin{eqnarray}
a^+ = n \cdot a, \qquad a^- = \bn \cdot a.\nonumber
\end{eqnarray}
The $\perp$ components satisfy $n \cdot \mathbf{a}_\perp=\bn  \cdot \mathbf{a}_\perp=v \cdot n \cdot \mathbf{a}_\perp=0$.

The $\epsilon$ tensor is defined so that $\epsilon^{0123}=-\epsilon_{0123}=1$, and
\begin{eqnarray}
\epsilon_\perp^{\mu \nu} &=& \epsilon^{\mu \nu\alpha\beta}v_\alpha n_\beta, \nn[3pt]
g_\perp^{\mu \nu} &=& g^{\alpha \beta} - (n^\alpha v^\beta + n^\beta v^\alpha) + n^\alpha n^\beta. \nonumber
\end{eqnarray}

The momentum of the $b$ quark in the $B$ rest frame is $p_b = m_b v+k$, where $k\sim \lqcd$ is the residual momentum, and the coordinate choice is such that $\mathbf{k}_\perp=0$. Let $q^2$ be the invariant mass of the virtual gauge boson; $q^2=0$ for $B \to X_s \gamma$.  The $b$ quark decays into a light quark with momentum $p$ and gauge boson with momentum $q$, $p_b=p+q$. The momentum components of the particles are:
\begin{eqnarray}
\begin{array}{rclrcl}
p_b^+ &=& m_b + k^+ ,&
p_b^- &=& m_b + k^- ,\\[5pt]
q^+ &=& m_b x ,&
q^- &=& {q^2 \over m_b x }, \\[5pt]
p^+ &=& m_b(1-x)+k^+ ,&
p^- &=& {m_b^2 x -q^2 \over m_b x} + k^-, \\[5pt]
\multicolumn{6}{c}{{\mathbf p}_{b \perp} = {\mathbf p}_{\perp} ={\mathbf q}_{\perp} = 0.}\\
\end{array}
\label{2.13}
\end{eqnarray}
For $B \to X_s \gamma$ decay, the momentum components are given by Eq.~(\ref{2.13}) for $q$ and $p$, with $q^2 \to 0$.

The kinematics simplifies in shape function region $ x \to 1$ with $m_b(1-x) \sim k^+ \sim \lqcd$. For $B \to X_s \gamma$,
\begin{eqnarray}
p^- &\sim& m_b,\nn
p^+ &\sim& m_b(1-x)+k^+, \nonumber
\end{eqnarray}
and for $B \to X e^+ e^-$ and $B \to X_u \ell \bar \nu$ decays,
\begin{eqnarray}
\begin{array}{rclrcl}
q^+ &=& m_b,&
q^- &=& {q^2 \over m_b}, \\[5pt]
p^+ &=& m_b(1-x)+k^+ ,&
p^- &=& {m_b^2 -q^2 \over m_b}.
\end{array} \nonumber
\end{eqnarray}

Instead of decay distributions as a function of $x$, one can also study their moments. For a function $f(x)$ defined on $x\in [0,1]$, the moments are defined by
\begin{eqnarray}
M_N(f) &=& \int_0^1 {\rm d}x\ x^{N-1} f(x).\nonumber
\end{eqnarray}
The endpoint region $x\to1$ corresponds to large moments $N \to \infty$ with the heuristic relation $1-x \sim 1/N$. For smeared observables in the endpoint region with smearing width $\Delta$, the relation is $N \sim 1/\Delta$. The shape function region corresponds to taking large moments with $m_b/N \sim \lqcd$. It is also convenient to define $\bN = N e^{\gamma_E}=1.78 N$. We will compute all terms which do not vanish in the limit $N \to \infty$. The vanishing terms can be obtained by computing to higher order in the SCET expansion parameter $\lambda$.

\section{Mass Scales}\label{sec:mass}

$B$ decays in the endpoint region have four important scales which are relevant for our computation. The largest scale in the calculation is of order the $b$ quark mass, and is chosen to be $\mu_1=m_b$ or $\mu_1=p^-=(m_b^2-q^2)/m_b$.  At this scale, $p^+$ and $\lqcd$ are infrared scales and can be neglected, so the final state quark can be treated as massless. The appropriate effective theory to use at this scale is SCET, where the quark is described by a $n$-collinear field. The QCD operators match onto SCET currents at $\mu_1$. Our calculation does not simultaneously minimize logarithms of $m_b/\mu$ and $p^-/\mu$, so the choice of $\mu_1$ is a matter of taste. $\mu_1=m_b$ has the advantage that it does not depend on the kinematic variable $q^2$. We will assume that $p^-$ is of order $m_b$, i.e.\ that $q^2 \not \approx m_b^2$.

The next important scale is the invariant mass of the final hadronic system $\mu_2=\sqrt{p^2}\equiv p=(m_b^2-q^2)(1-x)$. Below this scale, the invariant mass of the final state hadrons is large, and they can be integrated out. At the scale $\mu_2$, one integrates out the final hadronic states by computing the time-ordered product of two SCET currents, integrating out the intermediate collinear quark, and matching on to bilocal heavy quark operators. The non-locality of the operators is set by  $\mu_3=p^+$. Below the next scale $\mu_3$, these non-local operators can be replaced by local operators. Finally, large logarithms in the matrix elements of the operators can be minimized by renormalizing them at the scale $\mu_4$ of order  the scale $\lqcd$ of non-perturbative dynamics. $\mu_4$ is chosen to be large enough that perturbation theory is still valid. The scales are summarized in Table~\ref{tab:mass}.
\begin{table}
\renewcommand{\arraystretch}{2}
\begin{eqnarray*}
\begin{array}{c|cc}
\hline\hline
\mu_1 & m_b,\ \bn \cdot p = {m_b^2 - q^2 \over m_b}  \\
\mu_2 & p = \sqrt{p^+ p^-} =\sqrt {(m_b^2-q^2)(1-x) } & =\sqrt{m_b^2 - q^2 \over  \bN } \\
\mu_3 & p^+  = m_b(1-x) &= {m_b \over \bN} \\
\mu_4 & \lqcd\\
\hline\hline
\end{array}
\end{eqnarray*}
\caption{The scales $\mu_1$--$\mu_4$ that we will use. $\mu_4$ is a scale of order $\lqcd$ at which perturbation theory  is still valid, $\mu_4 \sim 1$~GeV.
\label{tab:mass}}
\end{table}

There is one important difference from deep inelastic scattering, the existence of the scale $\mu_3$. Boost invariance in deep inelastic scattering forbids the occurrence of such a scale. However, in the case of heavy meson decays, boost invariance is broken by the choice of rest frame of the $B$ meson. Equivalently, the decay amplitudes can depend on $v$, which does not exist for deep inelastic scattering, and so one can have the scale $p^+$.

Finally, keep in mind that the ratio of the largest and smallest scales $\mu_1/\mu_4$ is at most five in $B$ decays.

\section{The Hadronic Tensor and the Endpoint region}\label{sec:htensor}

All strong interaction effects for the inclusive $B$ decays studied here can be encoded in the hadronic tensor
\begin{eqnarray}
W^{\alpha\beta} = - \frac{1}{\pi} {\rm Im}\ T^{\alpha\beta},\nonumber
\end{eqnarray}
where
\begin{eqnarray}\label{Tdef}
T_{(f)}^{\alpha\beta} &=& -i \int d^4 x e^{-i q \cdot x} \frac{\left \langle \bar B \left| T\left[J_{(f)}^{\dagger\alpha}(x)  J_{(f)}^\beta(0) \right] \right| \bar B \right\rangle}{2m_B},\nonumber
\end{eqnarray}
and $J^\alpha_{(u)}$ and  $J^\alpha_{(s)}$ are the quark currents mediating the $b \to u \ell \bar \nu$ and $b \to s \gamma$ transition, respectively,
\begin{eqnarray}
J_{(s)} ^\alpha &=& \frac{1}{m_b}\bar s P_R\, \sigma^{\alpha\beta} \, b \,\,q_\beta, \nn
J_{(u)}^\alpha &=& \bar u P_R \, \gamma^\alpha \, b .\nonumber
\end{eqnarray}
A factor of $1/m_b$ has been included in the $b \to s \gamma$ current so that it has the same dimension as the $b \to u$ current.

The hadronic tensor $W^{\alpha\beta}$ can only depend on the momenta $p_b$ and $q$, or equivalently, on the velocity of the heavy quark, $v^\mu$ and the momentum of the light quark $p^\mu$. The most general tensor structure possible is
\begin{eqnarray}
W^{(u,s)\alpha\beta}& =  &
-g^{\alpha\beta} W^{(u,s)}_1 
+  v^\alpha v^\beta W^{(u,s)}_2 
+ i \epsilon^{\alpha\beta\rho\sigma} v_\rho q_\sigma W^{(u,s)}_3\nn
&& + q^\alpha q^\beta W^{(u,s)}_4
+ \left( q^\alpha v^\beta + q^\beta v^\alpha \right) W^{(u,s)}_5.
\label{Wdecomp}\end{eqnarray}
using the convention of Refs.~\cite{MW,book} for $W_i$. The coefficient of the $W_3$ term is the opposite of Refs.~\cite{MW,book} because we use the opposite sign convention for $\epsilon_{0123}$. The scalar functions $W_i^{(f)}$ depend on all possible Lorentz invariants that can be formed from the two vectors $v^\mu$ and $p^\mu$. There are two such invariants, and we will chose them to be $\bn \cdot p$ and $n \cdot p$,
\begin{eqnarray}
W^{(f)}_i \equiv W^{(f)}_i(\bn \cdot p, n \cdot p).\nonumber
\end{eqnarray}

The inclusive differential decay rates for the decays $B\to X_s \gamma$ and $B \to X_u \ell \bar\nu$ written in terms of the scalar functions $W^{(f)}_i(n \cdot p, \bn \cdot p)$ are
\begin{eqnarray}
\frac{{\rm d} \Gamma^s}{{\rm d}x_\gamma} &=& 2\, m_b\, \Gamma_0^{(s)}  x_\gamma \left[ 4W_1^{(s)}- W_2^{(s)} - x_\gamma\, m_b \,  W_5^{(s)} \right],\nn
\frac{{\rm d} \Gamma^u}{{\rm d}x_\ell\, {\rm d} z\, {\rm d} \hat p^2}&=& 12\, m_b\, \Gamma_0^{(u)} \Bigl[
2 (1-z+\hat p^2) W_1^{(u)} \nn
&& + (\bar x_\ell (z-\bar x_\ell)-\hat p^2) W_2^{(u)} \nn
&& + m_b(1-z+\hat p^2)(z-2\bar x_\ell) W_3^{(u)} \Bigr],
\label{diffrate}
\end{eqnarray}
where
\begin{eqnarray}
\Gamma_0^{(s)} &=& \frac{G_F^2 \left|V_{tb}V_{ts}\right|^2 \alpha \left| c_7^{\rm eff} \right|^2m_b^5}{32 \pi^4} ,\nn
\Gamma_0^{(u)} &=& \frac{G_F^2 \left| V_{ub} \right|^2 m_b^5}{192 \pi^3},\nonumber
\end{eqnarray}
and we have defined the dimensionless variables
\begin{eqnarray}
x_\gamma = \frac{2 E_\gamma}{m_b}\,, \quad x_\ell = \frac{2 E_\ell}{m_b}\,, \quad \hat p^2 = \frac{p^2}{m_b^2}\,, \quad z = \frac{2 v\cdot p}{m_b}\,,\nonumber
\end{eqnarray}
and $\bar x_\ell = 1-x_\ell$.  $c_7^{\rm eff}$ is the coefficient of the $b \to s \gamma$ operator $O_7$ in the weak Hamiltonian at the scale $\mu_1$. At next-to-leading order $c_7(m_b)=-0.311$ \cite{c7}.

The decay $B \to X_s \ell^+ \ell^-$ can be treated in a similar way, however one has to take into account the presence of left and right handed fermions. The expressions of the decay rates in this case can be obtained, for example, from Ref.~\cite{bsgOPE}.

The hadronic tensor can be calculated using an operator product expansion for the time-ordered product of the two currents in Eq.~(\ref{Tdef}). This procedure has been explained in great detail in \cite{MW} and the scalar functions $W_i$ are known perturbatively to order $\alpha_s$ \cite{dFN} and non-perturbatively to order $1/m_b^3$~\cite{3m}. For these results to be applicable, the phase space for the decay has to be dominated by a region where the invariant mass of the final hadronic system is far away from zero. Unfortunately, for many experimentally accessible observables this condition is not satisfied. 

In the endpoint region, where $p^- \gg p^+$, or equivalently $E_p^2 \gg p^2$, the traditional OPE breaks down. However an operator product expansion in terms of bilocal operators is still possible. The leading order operator (in the SCET expansion parameter $\lambda\sim \sqrt{p^+/p^-}$) is the bilocal operator
\begin{eqnarray}
O_v(k^+)&=&{1\over 2 \pi}\int_{-\infty}^\infty dx^- e^{-i x^- k^+} \bar b_v(0) \,Y(0,x^-)\,b_v(x^-) \nn
&=& \bar b_v \delta(in \cdot D + k^+) b_v, 
\label{Ovdef}
\end{eqnarray}
where $Y$ is an eikonal Wilson line in the $n$ direction from $x^-$ to $0$.The scalar functions $W^{(f)}_i(\bn \cdot p, n \cdot p)$ can be written as convolutions of perturbatively calculable Wilson coefficients with the matrix element of this operator
\begin{eqnarray}
W^{(f)}_i(\bn \cdot p, n \cdot p) = \! \int \! dk^+ C^{(f)}_i(\bn \cdot p,p^+- k^+,\mu) f(k^+,\mu),\nn
\label{Wieff}
\end{eqnarray}
where
\begin{eqnarray}
f(k^+,\mu) = \frac{\langle \bar B | O_v(k^+,\mu) | \bar B \rangle}{2m_B}.
\label{fdef}
\end{eqnarray}
The lightcone distribution function (shape function) of the $B$ meson, $f(k^+)$, encodes all the non-perturbative effects of the inclusive decay at leading order in the SCET expansion parameter $\lambda$, and has to be determined from experiment. It satisfies
\begin{eqnarray}
\label{fvanish}
f(k^+,\mu) = 0 \quad \mbox{for} \, \,\,k^+ < -\bar \Lambda
\end{eqnarray}
since the matrix element $\me{B}{O_v(k^+,\mu)}{B}$ has no discontinuities for $k^+ < -\bar \Lambda$. All the remaining information of the inclusive decay rates is encoded in the perturbative Wilson coefficients $C^{(f)}_i$. 

As explained above, there are four relevant scales in the problem. The region of phase space considered here and for which Eq.~(\ref{Wieff}) is valid requires $\mu_1 \gg \mu_2 \gg \lqcd$. The calculation proceeds as follows: (1) Match from QCD to SCET currents at $\mu_1$. (2) Run from $\mu_1$ to $\mu_2$ using the SCET anomalous dimension $\gamma_1$. (3) Integrate out the final hadronic states at $\mu_2$ by computing the time-ordered product of currents, and match onto bilocal operators (4) Run the bilocal 
operators to some common scale $\mu\sim 1$~GeV. The matrix element of the bilocal operator is the non-perturbative shape function $f(k^+,\mu)$ and Eq.~(\ref{Wieff}) is the final result. 

However, if $\mu_3 \gg \lqcd$, one can proceed further. The running in step (4) is then performed down to the scale $\mu_3$ and then (5) Match onto local operators at the scale $\mu_3$ (6) Run the local operators down to the scale $\mu_4$ and compute matrix elements. The matrix elements of these local operators are the usual HQET parameters $\lambda_i$, $\rho_i$ etc.\ which also occur in the OPE for totally inclusive processes.

Steps (1)--(4) will be performed in section \ref{sec:pert}, while the steps (5)--(6) will be performed in section \ref{sec:local}. 

\section{$W^{(f)}_i$ in the shape function region $p^+ \sim \lqcd$.}
\label{sec:pert}

\subsection{Matching from QCD to SCET}
\label{sect_matchcurrent}

The one-loop matching condition between QCD and SCET is given by computing the on-shell matrix element in QCD using dimensional regularization to regulate both the ultraviolet and infrared divergences, and keeping only the finite part~\cite{AM}.  The required calculations  have been performed in \cite{BFPS} and we will just collect the results given there. 

Since both the light quark field as well as the heavy quark field are given by two component spinors in the effective theory, only three different Dirac structures are possible for heavy to light currents in the effective theory
\begin{eqnarray}
j^{\rm eff} = \bar \chi_n \, \Gamma \, b_v, \qquad {\rm with} 
\qquad \Gamma = 1, \quad \gamma^5, \quad \gamma_\perp^\mu\,, \nonumber
\end{eqnarray}
where $\chi_n$ is the collinear light quark field.

In order to facilitate writing left and right handed currents, we will often write a fourth Dirac structure, which however is related to the previous three
\begin{eqnarray}
\gamma_\perp^\mu \gamma^5 \equiv \epsilon_\perp^{\mu \nu} \gamma^\perp_\nu, \nonumber
\end{eqnarray}
where, as before, $\epsilon_\perp^{\mu \nu} = \epsilon^{\mu \nu\alpha\beta}v_\alpha n_\beta$ with $\epsilon_{0123}=-1$. 
The collinear field is defined as
\begin{eqnarray}
\chi_n(x) = \sum_{\tilde p} e^{-i \tilde p \cdot x} [W_n \xi_{n}](x), \nonumber
\end{eqnarray}
where $\tilde p$ is the label momentum which contains components of order $1$ and order $\lambda$. The order $\lambda^2$ components are associated with the spacetime dependence of the fields. 
$W_n(x)$ denotes a Wilson line of collinear gluons along the path in the $n$ direction from $x$ to $\infty$ . This Wilson line is required to ensure gauge invariance of the current in the effective theory \cite{BS}. 

Since for the processes we are considering we only need left handed light quarks, there are only two currents in SCET. Define
\begin{eqnarray}
j_{1\mu}^{\rm eff} &=& \bar \chi_{n}P_R\, \gamma_\mu^\perp b_v\,,\nn
j_2^{\rm eff} &=& \bar \chi_{n}\, P_R \,b_v\,.\nonumber
\end{eqnarray}

The full theory currents mediating the decays $B \to X_s \gamma$ and $B \to X_u \ell \bar \nu$ can be matched onto these effective theory currents. Using the results of \cite{BFPS} we find
\begin{widetext}
\begin{eqnarray} \label{ctrnsfm}
\bar uP_R \gamma_\mu b &\to& C_1(\bn \cdot p, \mu_1)\, j_{1\mu}^{\rm eff} +
   \Big\{ C_2(\bn \cdot p, \mu_1)\, n_\mu +C_3(\bn \cdot p, \mu_1)\, v_\mu\Big\} j_2^{\rm eff}
   \,,\nn
\bar sP_R \, i\sigma_{\mu\nu} b &\to& 
  C_4(\bn \cdot p, \mu_1)\,( n_\mu g_{\nu\lambda}\! -\!  n_\nu g_{\mu\lambda})
  j_{1}^{\rm eff\, \lambda}
+ C_{5}(\bn \cdot p, \mu_1)\, (v_\mu n_\nu -  v_\nu n_\mu + i \epsilon^\perp_{\mu\nu})\, j_2^{\rm eff} \,.
\end{eqnarray}
The Wilson coefficients $C_{1-5}$ are given by
\begin{eqnarray} \label{matchC}
C_1(\bn \cdot p, m_b, \mu) &=& 1 - \frac{\alpha_s(\mu)C_F}{4\pi} \Bigg\{ 
g(\bn \cdot p, m_b, \mu) + \ln(\bn \cdot \hat p) \left( \frac{3\bn \cdot \hat p-2}{1-\bn \cdot \hat p}\right) \Bigg\} , \nn 
C_2(\bn \cdot p, m_b, \mu)  &=& 1 - \frac{\alpha_s(\mu)C_F}{4\pi} \Bigg\{g(\bn \cdot p, m_b, \mu)  - \ln(\bn \cdot \hat p) \bigg[ \frac{2-4\bn \cdot \hat p+(\bn \cdot \hat p)^2}
  {(1-\bn \cdot \hat p)^2}\bigg] +\frac{\bn \cdot \hat p}{1-\bn \cdot \hat p} \Bigg\} 
 \,, \nn 
C_3(\bn \cdot p, m_b,\mu)  &=& \frac{\alpha_s(\mu)C_F}{4\pi}\: \Bigg\{ \frac{2}{(1-\bn \cdot \hat p)} 
  + \frac{2\bn \cdot \hat p\ln(\bn \cdot \hat p)} {(1-\bn \cdot \hat p)^2} \Bigg\} \,,\\
C_{4}(\bn \cdot p, m_b, \mu)  &=& 1 - \frac{\alpha_s(\mu)C_F}{4\pi} \Bigg\{
g(\bn \cdot p, m_b, \mu) + \ln {\mu^2 \over m_b^2} -2 \ln(\bn \cdot \hat p)  \Bigg\}
  \,,\nn
C_{5}(\bn \cdot p, m_b, \mu)  &=& 1 - \frac{\alpha_s(\mu)C_F}{4\pi} \Biggl\{g(\bn \cdot p, m_b, \mu)  + \ln {\mu^2 \over m_b^2} +  \ln(\bn \cdot \hat p) \left( \frac{4\bn \cdot \hat p-2}{1-\bn \cdot \hat p}\right) \Biggr\} , \nonumber\,
  \end{eqnarray}
where 
\begin{eqnarray}
g(\bn \cdot p, m_b, \mu) &=& \frac 1 2 \ln^2{ \mu^2 \over m_b^2} + \frac 5 2 \ln {\mu^2 \over m_b^2} -
2 \ln {\mu^2 \over m_b^2}\ln(\bn \cdot \hat p) + 2\!\ln^2(\bn \cdot \hat p) 
  + 2 {\rm Li}_2(1\!-\!\bn \cdot \hat p)   + \frac{\pi^2}{12} + 6 \,, \nonumber
\end{eqnarray}
\end{widetext}
and
\begin{eqnarray}
\bn \cdot \hat p = {\bn \cdot p \over m_b} &=& 1 - {q^2 \over m_b^2}. \nonumber
\end{eqnarray} 
The matching coefficients for $\sigma^{\mu \nu}$ have an extra $\ln \mu^2/m_b^2$ contribution which corresponds to the anomalous dimension for this operator in full QCD. 

The matching scale can be chosen to be either $\mu_1=\bn \cdot p$ or $\mu_1 =m_b$ At these scales the matching coefficients are given by Eq.~(\ref{matchC}) with
\begin{eqnarray} 
g(\bn \cdot p, m_b, \bn \cdot p) &=& 5 \ln( \bn \cdot \hat p) 
  + 2 {\rm Li}_2(1\!-\!\bn \cdot \hat p)   + \frac{\pi^2}{12} + 6,\nn
g(\bn \cdot p, m_b, m_b) &=& 2\ln^2(\bn \cdot \hat p)
  + 2 {\rm Li}_2(1\!-\!\bn \cdot \hat p)   + \frac{\pi^2}{12} + 6 .\nonumber
\end{eqnarray}

\subsection{Renormalization of the SCET Currents}

The running of the currents in the effective theory have also been calculated in \cite{BFPS}. The currents are multiplicatively renormalized, and satisfy the renormalization group equation
\begin{eqnarray} 
  \mu \frac{{\rm d}}{{\rm d}\mu} C_i(\mu,\bn \cdot p) = \gamma(\mu,\bn \cdot p) C_i(\mu,\bn \cdot p)\,. \nonumber
\end{eqnarray}
The leading order (LO) and next to leading order (NLO) anomalous dimensions are 
\begin{eqnarray}
 \gamma_{\text{LO}} &=& -\frac{\alpha_s(\mu) C_F}{\pi} 
  \ln\bigg( \frac{\mu}{\mu_1} \bigg) \,, \nn
\gamma_{\text{NLO}} &=& -\frac{\alpha_s(\mu) C_F}{2 \pi} \left[\frac{5}{2}+2\ln \left(\frac{\mu_1}{\bn \cdot p}\right)\right]\nn
   && \hspace{.5cm} -2 B  \frac{\alpha_s^2(\mu)C_F }{(2\pi)^2}
  \,\ln\bigg( \frac{\mu}{\mu_1}\bigg) \,. 
 \label{adim2}\end{eqnarray}
where \cite{Bvalue}
\begin{eqnarray}
B =  C_A\left({67 \over 18}-{\pi^2\over6}\right)-{5 \over 9} n_f. \label{Bdef}
\end{eqnarray}
Note that the value of $\mu_1$ in these anomalous dimensions is chosen to coincide with the choice made for the matching scale in Eq.~(\ref{matchC}). The $\mu_1$ dependence in the $\alpha_s$ term cancels between $\gamma_{\text{LO}}$ and $\gamma_{\text{NLO}}$. The difference between the two possible choices for $\mu_1$ in the $\alpha_s^2$ term is NNLO.

The solution to this renormalization group equation is given by
\begin{eqnarray}
   C_i(\mu) &=& C_i(\mu_1) \exp\Bigg[\frac{r_0(z)}{\alpha_s(\mu_1)} 
  + r_1(z) \bigg]  \,,\nonumber
\end{eqnarray}
with
\begin{eqnarray}
r_0(z) &=& -\frac{4\pi C_F}{\beta_0^2}
 \:\Big[ \frac{1}{z} -1 + \ln z \Big] \,,\nn
r_1(z) &=& -\frac{ C_F\beta_1}{\beta_0^3} \Big[ 1 -z + \ln z 
-\frac12 \ln^2 z
  \Big] 
  \nn&&
+ \frac{C_F}{\beta_0} \Big[ \frac52 + 2 \ln\left(\frac{\mu_1}{\bn \cdot p}\right) \Big] \ln z  
  \nn
  &&
  - \frac{2 C_F B}{\beta_0^2}  \Big[ z -1- \ln z \Big]\,,
\label{fdef2}\end{eqnarray}
where
\begin{eqnarray}
  z &=& {\alpha_s(\mu)\over \alpha_s(\mu_1)} , \nn
  \beta_0 &=& \frac{11 }{3} C_A- \frac{2}{3} n_f ,\nn
  \beta_1 &=& \frac{34}{3}C_A^2-\frac{10}{3} C_A n_f-2 C_F n_f,\nn
\mu \frac{{\rm d}}{{\rm d}\mu} \alpha_s &=& -{\beta_0 \over 2 \pi} \alpha_s^2 - {\beta_1 \over 8 \pi^2} \alpha_s^3,\nonumber
 \end{eqnarray}
$C_A=3$, $C_F=4/3$ and $n_f=4$ is the number of light quark flavors.

\subsection{Matching to bilocal Operators}

\begin{figure}
\includegraphics[width=4cm]{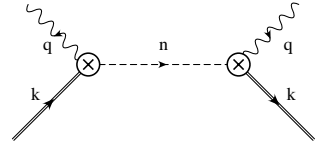}
\caption{Tree graph for the product of two currents. The dashed line depicts a collinear particle.\label{fig1}}
\end{figure}

\begin{figure}
\includegraphics[width=4cm]{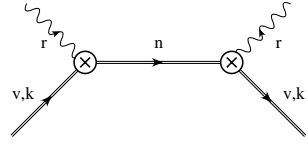}
\caption{Tree level matrix element of the shape function operator. The double line between the two currents depicts a Wilson line in the $n$ direction.\label{fig2}}
\end{figure}

Below the scale $\mu_2=p$ the final hadronic state is heavy and can be integrated out. This is done by matching the time-ordered product of two SCET currents onto bilocal operators by integrating out the $n$-collinear fields at the scale $\mu_2$. The tree level time-ordered product graph is shown in Fig.~\ref{fig1}. The tree level graph for the bilocal operator is shown in Fig.~\ref{fig2} and matches on to a bilocal heavy quark operator, the shape function operator defined in Eq.~(\ref{Ovdef}). The Feynman rules for matrix elements of this operator are given by taking the discontinuity of the diagram, since the operator is a product, not a time-ordered product.
Spinors in SCET have only two components, which can be used to simplify the results significantly.  One can easily show that after taking matrix elements between pseudoscalar $B$ mesons that the only non-vanishing time-ordered products have two identical currents $j_i^{\text{eff}}$.  Furthermore, perturbative corrections in SCET are independent of  the Dirac structure of the currents. As a result, one can show that the only non-vanishing time-ordered products in SCET can be written in the form
\begin{eqnarray}
&&-i \int d^4 x e^{-i q \cdot x} \left \langle \bar B \left| T\left[j_{1\alpha}^{{\rm eff}\dagger}(x)  \, , \, j_{1\beta}^{{\rm eff}}(0) \right] \right| \bar B \right\rangle 
\nn
&&\hspace{1cm}
= 
-\frac{1}{4} (g^\perp_{\alpha\beta} + i \epsilon^\perp_{\alpha\beta})\left \langle \bar B \left|  {\cal T}(p^+)\right| \bar B \right\rangle,\nn
&&-i \int d^4 x e^{-i q \cdot x} \left \langle \bar B \left| T\left[j_{2}^{{\rm eff}\dagger}(x) \, , \,  j_{2}^{{\rm eff}}(0) \right] \right| \bar B \right\rangle 
\nn
&&\hspace{1cm}
=
\frac{1}{4}\left \langle \bar B \left| {\cal T}(p^+)\right| \bar B \right\rangle. \nonumber
\end{eqnarray}
Combining this result with Eqs.~(\ref{Wdecomp}) and (\ref{ctrnsfm}) we find for the decay $B \to X_u \ell \bar \nu$
\begin{eqnarray}
W_1^{(u)} &=& \frac{C_1^2}{4}{ \left \langle \bar B \left| {\cal W}(p^+)\right| \bar B\right\rangle\over 2m_B} ,\nn
W_2^{(u)} &=& \left[\frac{\bn \cdot \hat p-1}{\bn \cdot \hat p^2} \, C_1^2 
+ \left(\frac{C_2}{\bn \cdot \hat p} + \frac{C_3}{2} \right)^2 \right]
{ \left \langle \bar B \left| {\cal W}(p^+)\right| \bar B\right\rangle\over 2m_B},\nn
W_3^{(u)} &=& \frac{1}{2m_b \bn \cdot \hat p} \, C_1^2{ \left \langle \bar B \left|  {\cal W}(p^+)\right| \bar B\right\rangle \over 2m_B} ,\nn
W_4^{(u)} &=& \frac{1}{m_b^2 (\bn \cdot \hat p)^2}(C_2^2 - C_1^2){\left \langle \bar B \left|  {\cal W}(p^+)\right| \bar B\right\rangle\over 2m_B}, \nn
W_5^{(u)} &=& \frac{1}{2m_b (\bn \cdot \hat p)^2} \Bigl[C_1^2 (2-\bn \cdot \hat p)-2 C_2^2 \nn
&&- C_2 C_3 \bn \cdot \hat p \Bigr]{\left \langle \bar B \left| {\cal W}(p^+)\right| \bar B\right\rangle\over 2m_B},
\label{Wmatchu}
\end{eqnarray}
where
\begin{eqnarray}
 {\cal W}(p^+) &=&-{1\over \pi} \, \text{Im}\  {\cal T}(p^+). \nonumber
\end{eqnarray}
For the decay $B \to X_s \gamma$ we find (using $q^2=0$)
\begin{eqnarray}
W_1^{(s)} &=& \frac{C_4^2}{4} {\left \langle \bar B \left|  {\cal W}(p^+)\right| \bar B\right\rangle\over 2m_B},\nn
 \qquad 
W_2^{(s)} &=& 0\,, \nn
\qquad 
W_3^{(s)} &=& \frac{C_4^2}{2m_b} {\left \langle \bar B \left|  {\cal W}(p^+)\right| \bar B\right\rangle\over 2m_B}, \nn
\qquad 
W_4^{(s)} &=& \frac{1}{4m_b^2}\left(C_5^2 - 4C_4^2\right) {\left \langle \bar B \left|  {\cal W}(p^+)\right| \bar B\right\rangle\over 2m_B}, \nn
\qquad 
W_5^{(s)} &=& \frac{C_4^2}{2m_b}{ \left \langle \bar B \left|  {\cal W}(p^+)\right| \bar B\right\rangle\over 2m_B}.
\label{Wmatchs}
\end{eqnarray}
Gauge invariance implies (using $q^2=0$) that $W_2^{(s)}=0$ and $2 W_1^{(s)} = m_b W_5^{(s)}$, so these relations are true to all orders in perturbation theory. 
For the photon spectrum in $B \to X_s \gamma$, we will need the linear combination
\begin{eqnarray}
W_\gamma^{(s)}=2 m_b( 4 W_1^{(s)} - W_2^{(s)} -m_b W_5^{(s)}),
\label{17.20}
\end{eqnarray}
obtained from Eq.~(\ref{diffrate}) with $x_\gamma \to 1$. The matching for this linear combination of $W_i^{(s)}$ is
\begin{eqnarray}
W_\gamma^{(s)} &=& m_b C_4^2 {\left \langle \bar B \left|  {\cal W}(p^+)\right| \bar B\right\rangle\over 2m_B}.
\label{Wmatchg}
\end{eqnarray}

\begin{figure*}[t!]
\def\size{5.75 cm}
\hbox{\vbox{\hbox to \size {\hfil \includegraphics[width=4cm]{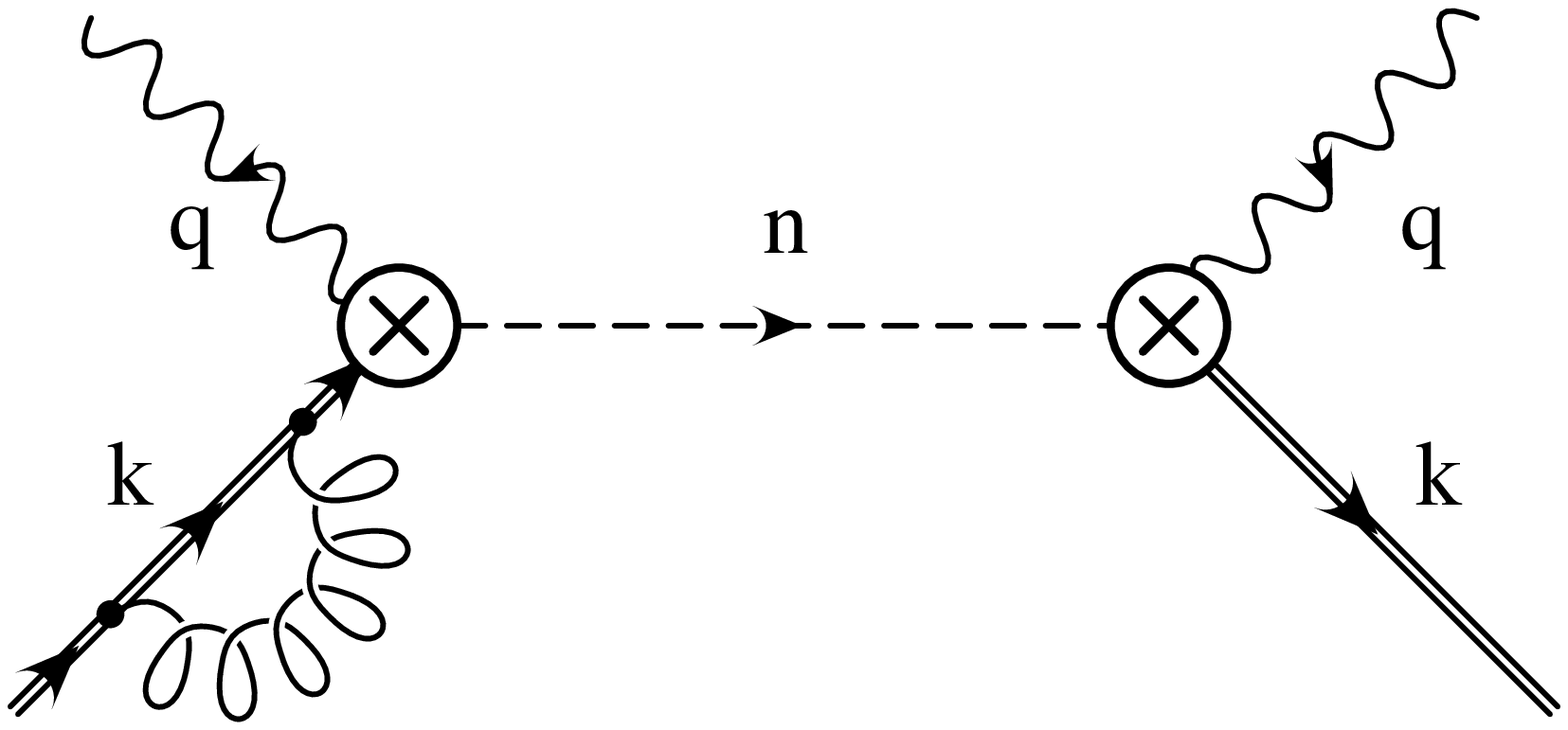} \hfil }\hbox to \size {\hfil(a)\hfil}}
\vbox{\hbox to \size {\hfil \includegraphics[width=4cm]{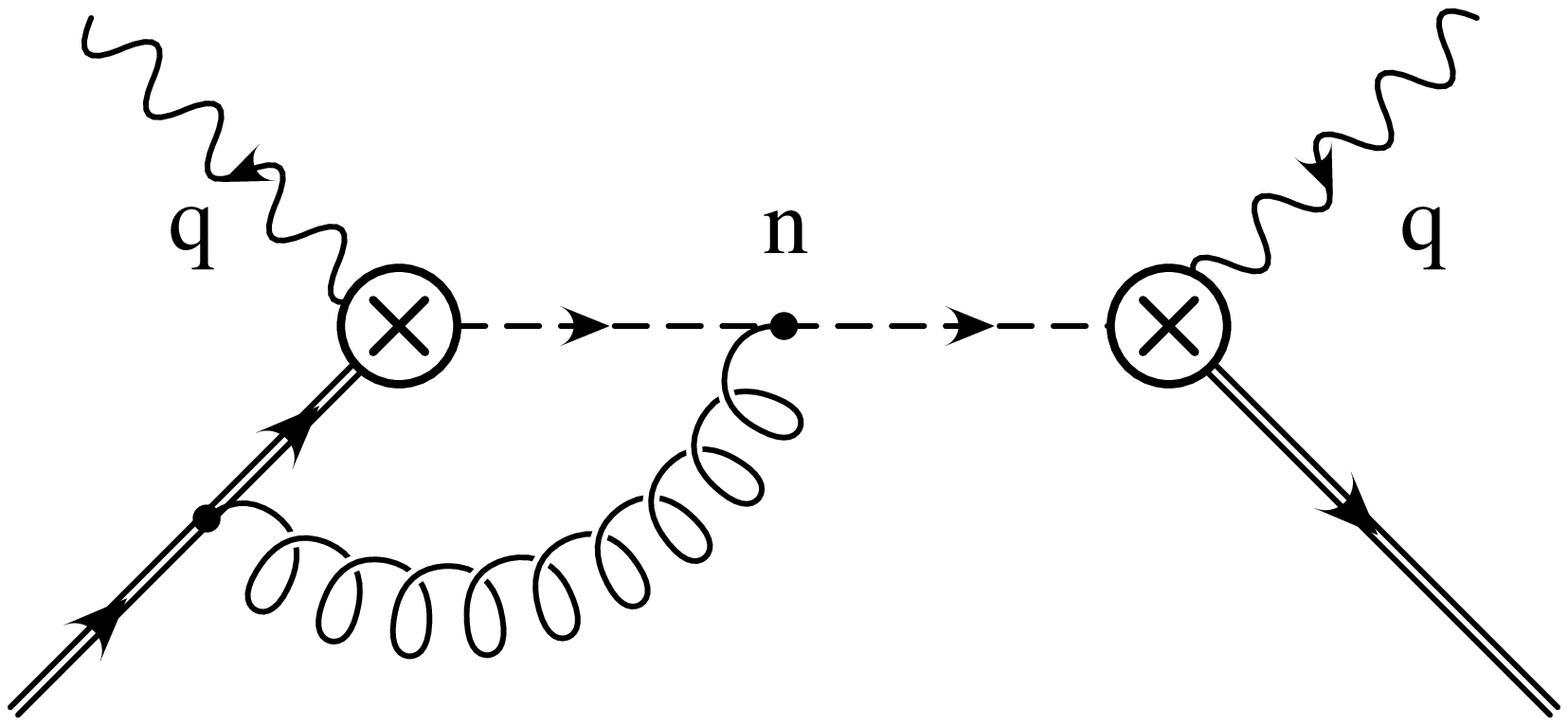} \hfil}\hbox to \size {\hfil(b)\hfil}}
\vbox{\hbox to \size {\hfil \includegraphics[width=4cm]{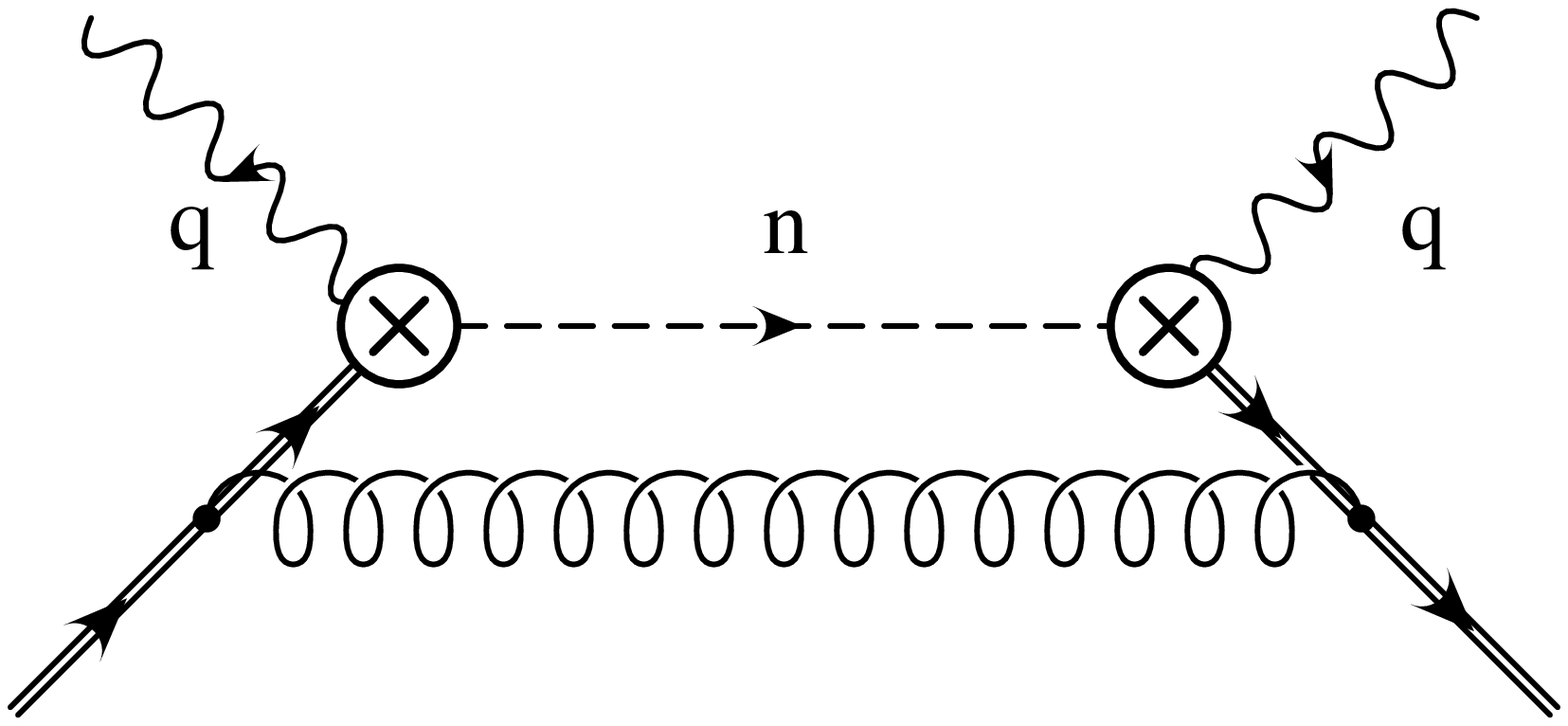} \hfil}\hbox to \size {\hfil(c)\hfil}}}
\hbox{\vbox{\hbox to \size {\hfil \includegraphics[width=4cm]{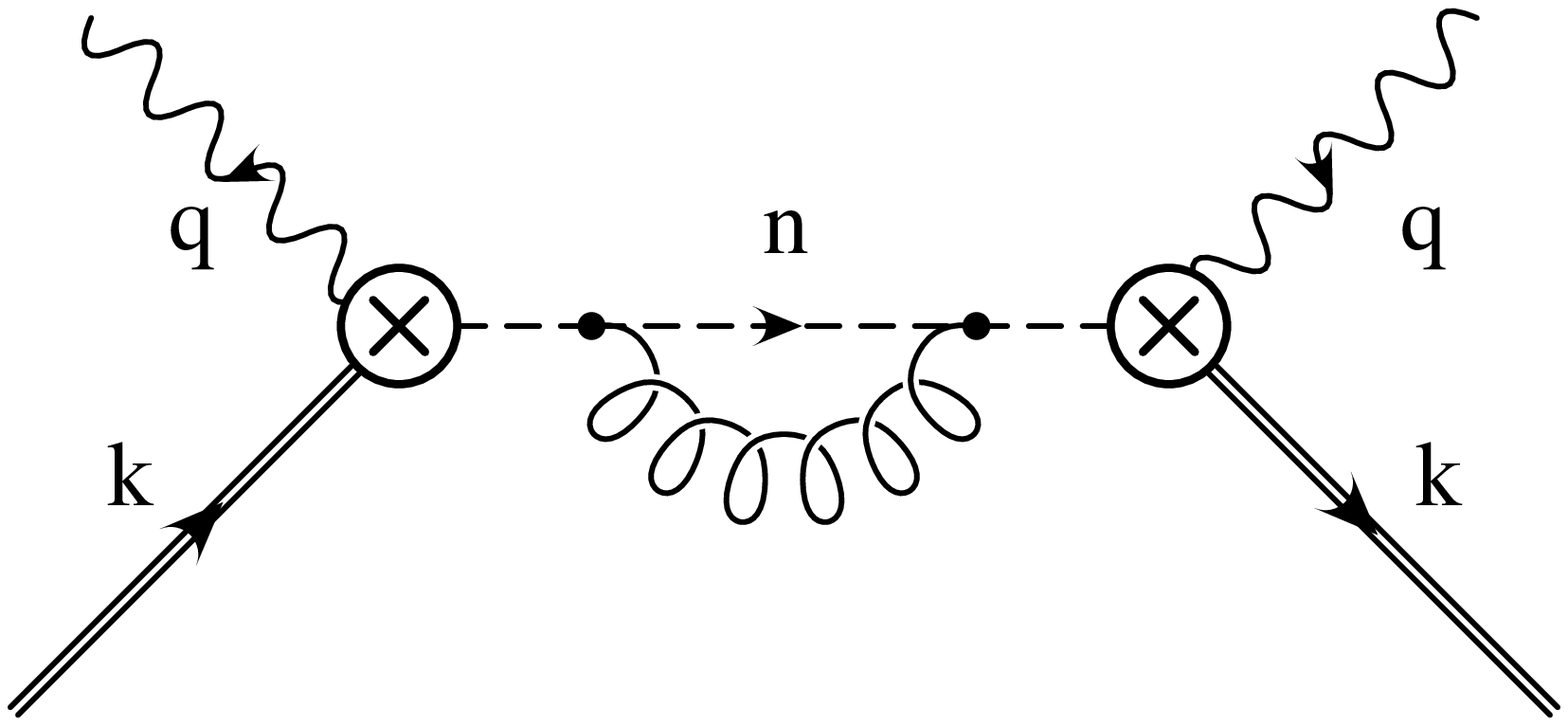} \hfil }\hbox to \size {\hfil(d)\hfil}}
\vbox{\hbox to \size {\hfil \includegraphics[width=4cm]{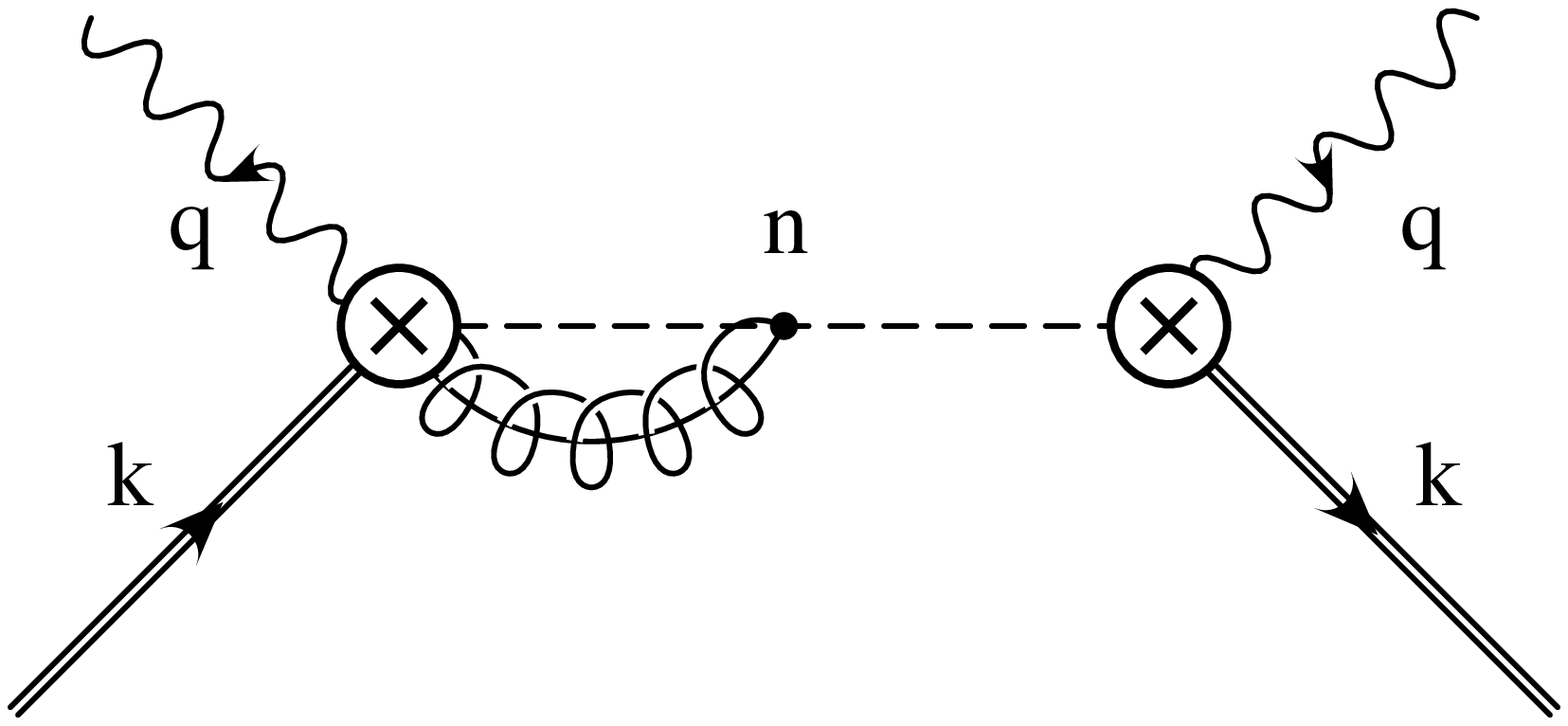} \hfil}\hbox to \size {\hfil(e)\hfil}}
\vbox{\hbox to \size {\hfil \includegraphics[width=4cm]{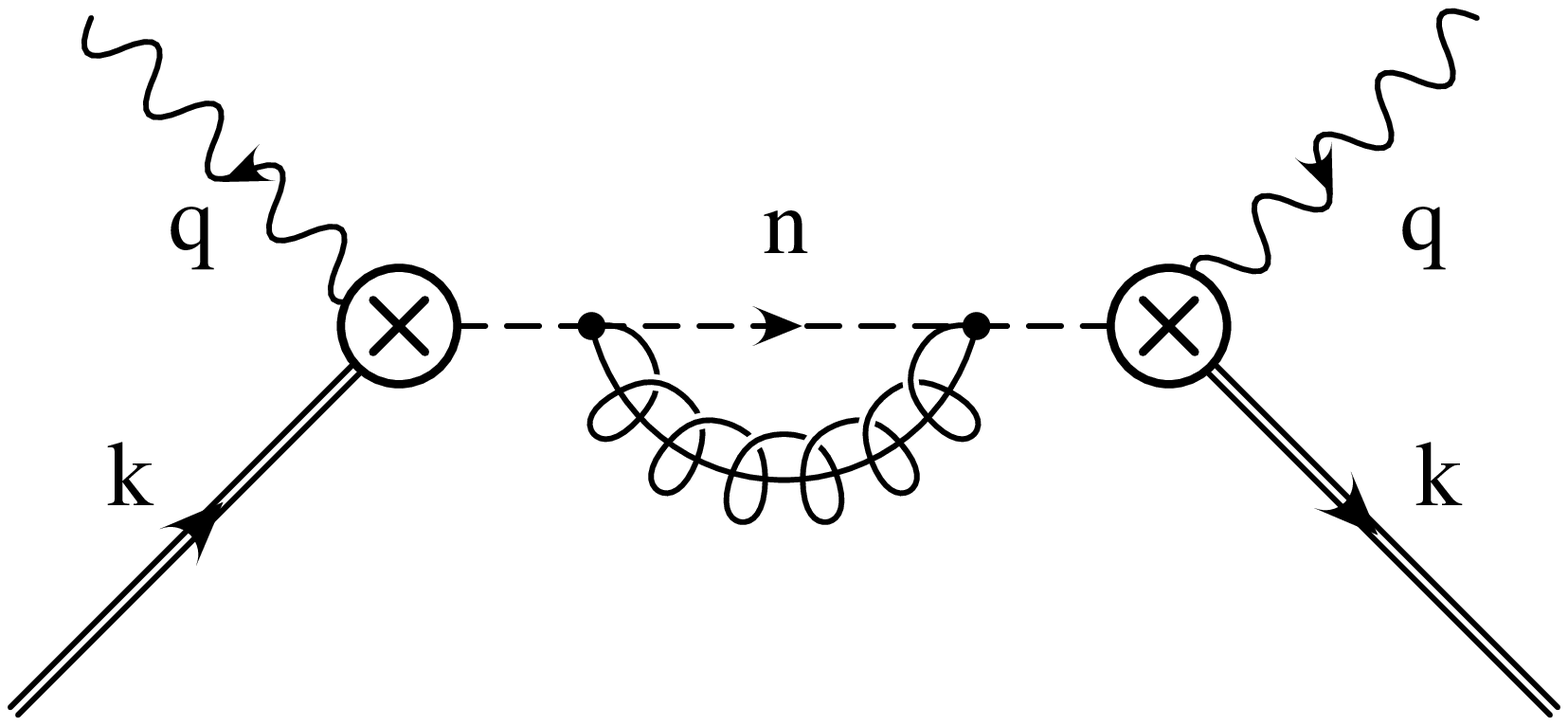} \hfil}\hbox to \size {\hfil(f)\hfil}}}
\hbox{\vbox{\hbox to \size {\hfil}}\vbox{\hbox to \size {\hfil \includegraphics[width=4cm]{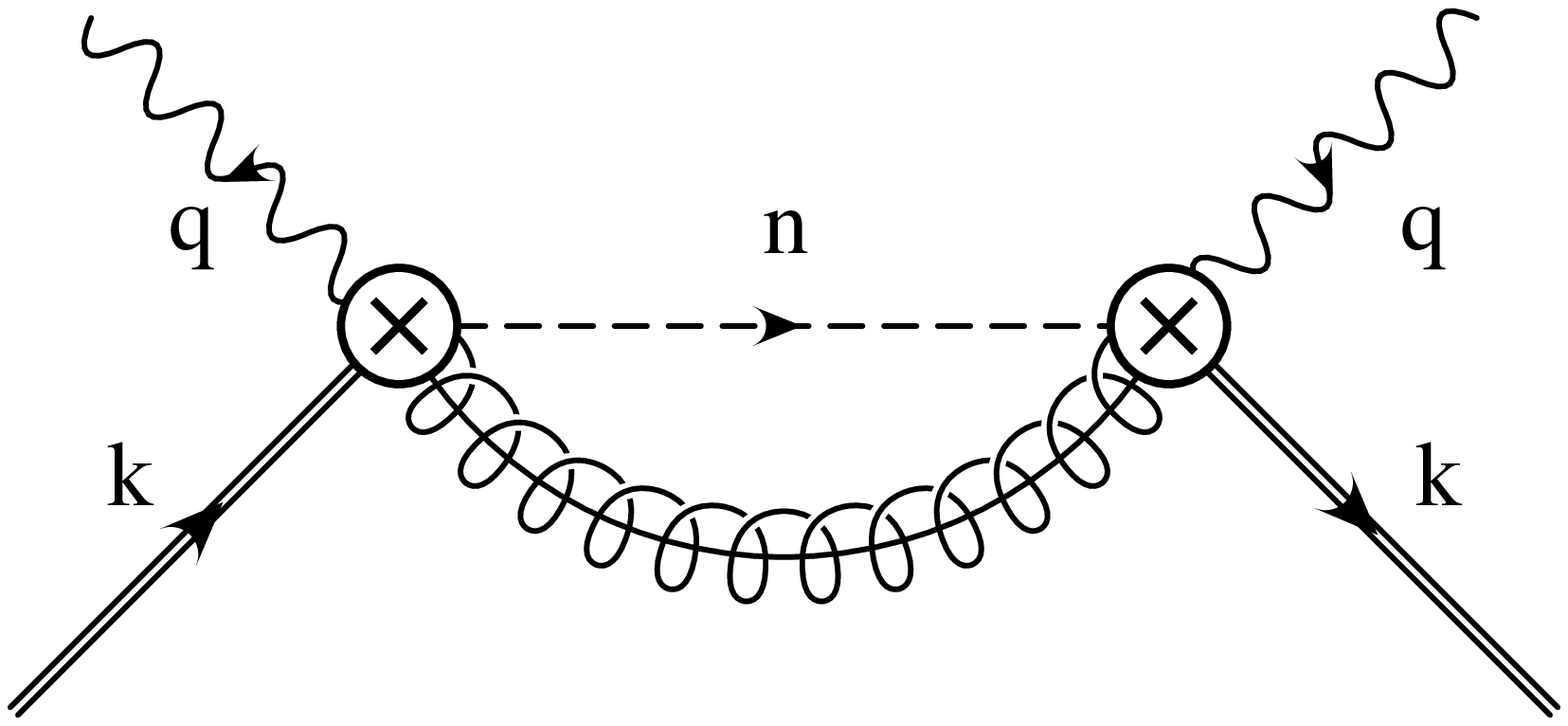} \hfil }\hbox to \size {\hfil(g)\hfil}}}
\caption{One loop correction to the current product. Graphs (a), (b) and (e) also have mirror image graphs where the gluon attaches to the other side. \label{fig3}}
\end{figure*}
\begin{figure}
\def\size{4.5 cm}
\hbox{\vbox{\hbox to \size {\hfil \includegraphics[width=4cm]{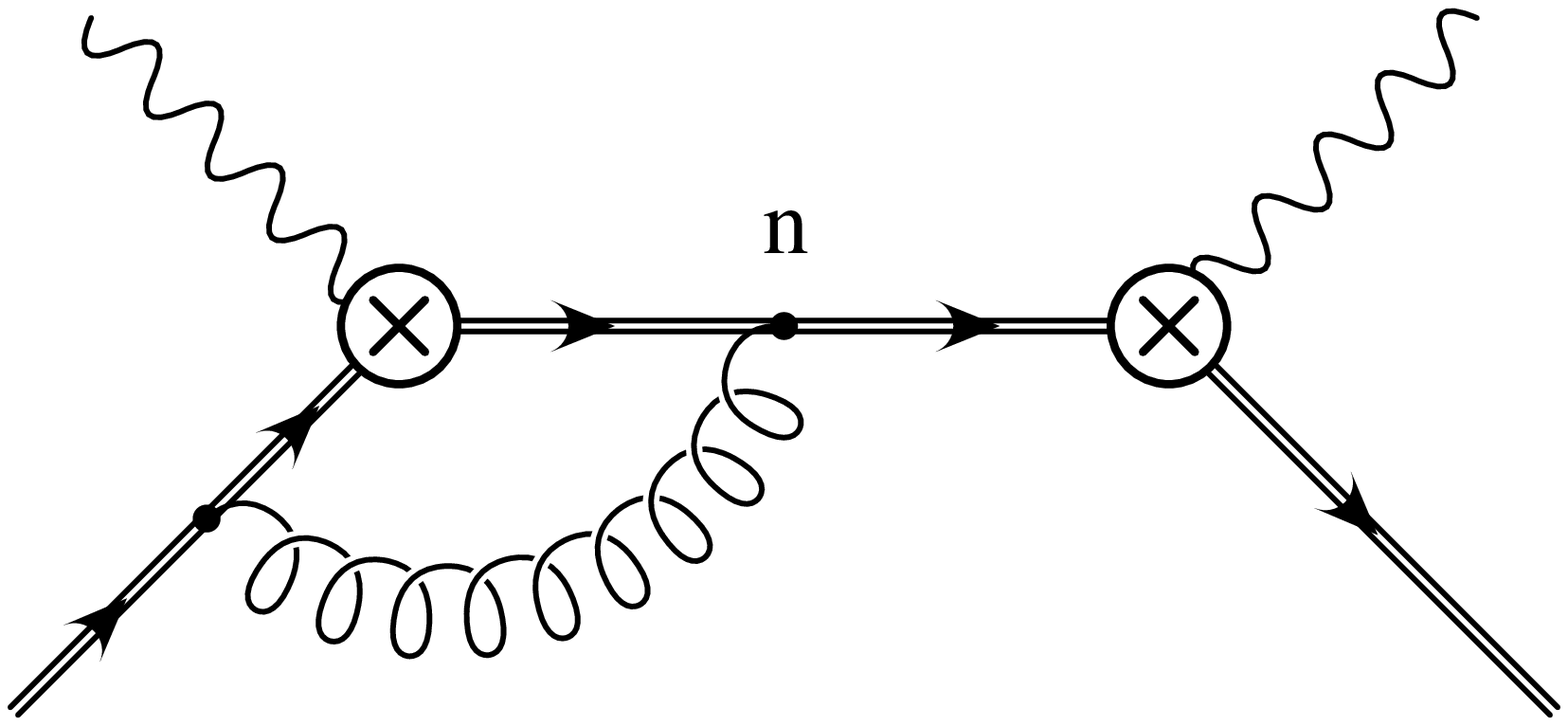} \hfil }\hbox to \size {\hfil(a)\hfil}}
\vbox{\hbox to \size {\hfil \includegraphics[width=4cm]{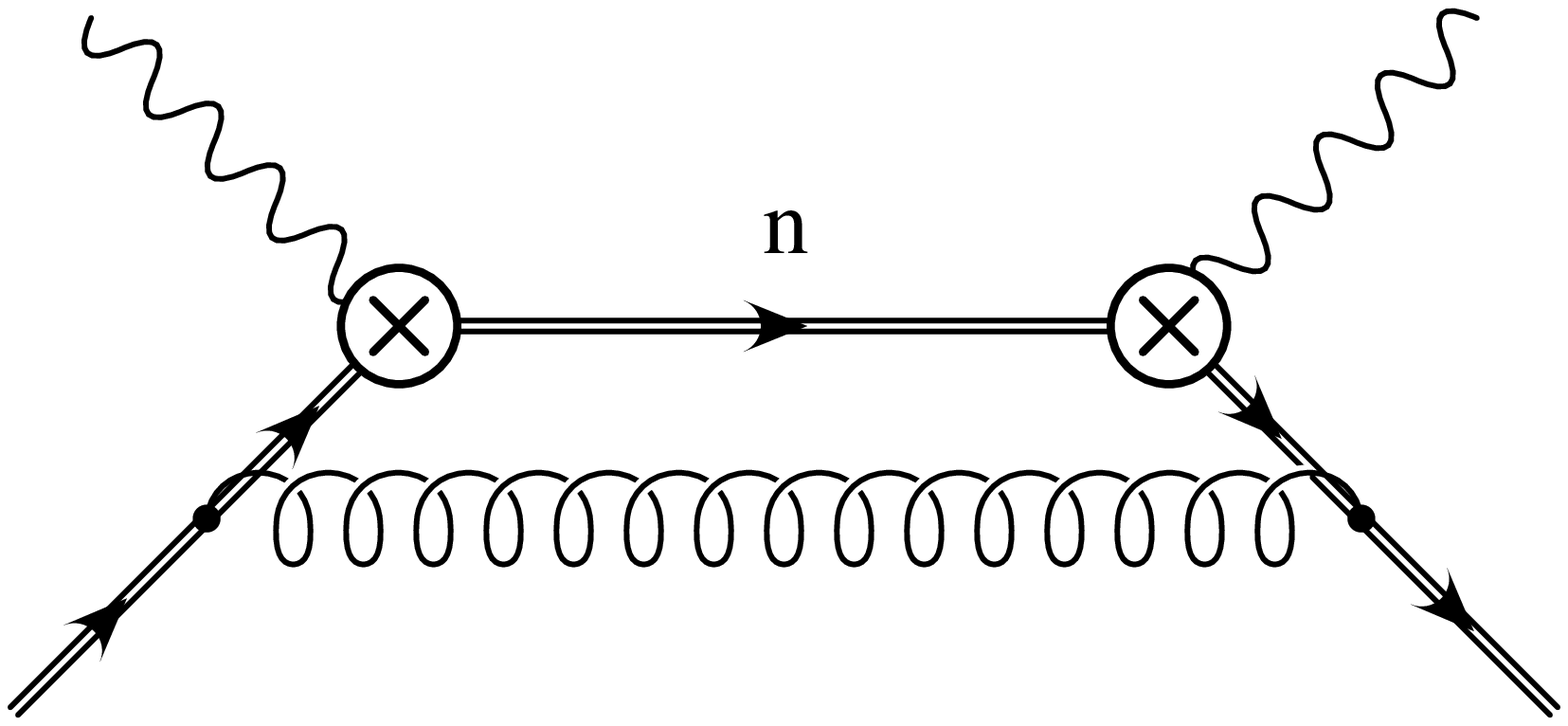} \hfil}\hbox to \size {\hfil(b)\hfil}}}
\caption{One loop correction to the shape function operator. \label{fig4}}
\end{figure}
Comparing the tree-level matrix element of the time-ordered product shown in Fig.~\ref{fig1} with the tree-level matrix element of the shape function operator shown in Fig.~\ref{fig2},
one finds
\begin{eqnarray}
{\cal W}(p^+) &=& O_v(p^+) + {\cal O}(\alpha_s). \nonumber
\end{eqnarray}

The one-loop matching condition is evaluated by computing the graphs in Fig.~\ref{fig3}, and subtracting from it the one-loop matrix element of the tree-level  operator ${\cal W}(p^+)$ in the effective theory, shown in Fig.~\ref{fig4}. The matrix element of $O_v(p^+)$ at one loop exactly reproduces the graphs in Fig.~\ref{fig3}(a--d), since the coupling of soft gluons to the $n$-collinear quark is identical to the coupling to the Wilson line $Y$. The matching condition is therefore given by graphs Fig.~\ref{fig3}(e--g). The Feynman rules for graphs Fig.~\ref{fig3}(e--g) do not depend on the external quark field, so the graphs have the same value as the matching graphs for deep inelastic scattering~\cite{dis}, with the replacement $p \to 0$, $q\to p$. The discontinuity of the graph extends over the infinite range $0 \le p^+ \le \infty$, rather than a finite interval such as $[0,1]$. For this reason, it is convenient to regulate the singularity at $p^+=0$ by a modified $+$-distribution, called the $\dist$-distribution since it depends on the scale $\mu$. The $\dist$-distribution, its properties, and relation to the $+$-distribution are given in Appendix~\ref{app:dist}. The resulting expression for ${\cal W}(p^+)$ is
\begin{eqnarray}
{\cal W}(p^+) &=& \int {{\rm d} r^+ }  \mathcal{C}(p^+-r^+,\mu) O_v(r^+) ,\nn
\mathcal{C}(q^+,\mu) &=& \mathcal{C} ^{(0)}(q^+,\mu) +{\alpha_s (\mu)C_F \over 4 \pi}\mathcal{C}^{(1)}(q^+,\mu) ,
\nn  \mathcal{C}^{(0)}(q^+,\mu) &=& \delta(q^+),\nn
\mathcal{C}^{(1)}(q^+,\mu) &=& 4  {\left[ \ln (q^+/\mu) \ \theta(q^+)\over q^+ \right]_{\dist} }  \nn
&& + \Bigl[4  \ln {\bn \cdot p \over  \mu} 
 -3  \Bigr]{ \left[\theta(q^+) \over q^+ \right]_{\dist} }\nn
 && \hspace{-1cm}+
 \Bigl[  2  \ln^2 {\bn \cdot p \over  \mu} 
- 3  \ln  {\bn \cdot p \over  \mu}+  7 - {\pi^2 } \Bigr] \delta(q^+) . \nn
\label{3.15}
\end{eqnarray}

\subsection{Renormalization of the Shape Function}

We can study the renormalization of the shape function by computing the on-shell matrix element of the shape function operator $O_v(r^+)$ in a heavy quark target, with residual momentum $k=0$.

The tree level matrix element is given by the graph shown in Fig.~\ref{fig2}. The spin-averaged matrix element is
\begin{eqnarray}
I_{\rm tree}&=&\text{Disc}{i\over 2 \pi} {1 \over r^+ + i 0^+}\nn
&=&  \delta\left(r^+ \right).
\label{4.03}
\end{eqnarray}

The loop corrections to the shape function matrix element are the graphs in Fig.~\ref{fig4} and wavefunction graphs.
Graph Fig.~\ref{fig4}(a) gives
\begin{eqnarray}
I_a
&=& - i  {g^2 C_F \over  16 \pi^2}\csc (2 \epsilon \pi) \Gamma(\epsilon) \mu^{2 \epsilon} \nn
&&\times\ \text{Disc}{ 1 \over  \left(r^+ + i 0^+ \right)} \left(-r^+  - i 0^+ \right)^{-2 \epsilon}\,. \nonumber
\end{eqnarray}
In terms of the $\dist$-distribution,
\begin{eqnarray}
I_a &= &-{g^2 C_F\over 16 \pi^2} \Biggl[
\left({1 \over \epsilon^2} +  { \pi^2 \over 12} \right) \delta(r^+)  \nn
&& -  {2 \over \epsilon}   \left[ \theta(r^+) \over r^+ \right]_{\dist} + 4 \left[ \ln (r^+/\mu) \ \theta(r^+) \over r^+ \right]_{\dist}\Biggr] .\nn
\label{4.14}
\end{eqnarray}

The spin-averaged loop graph Fig.~\ref{fig4}(b) is
\begin{eqnarray}
I_b  &=& i{g^2 C_F \over 16 \pi^3}\text{Disc}\left( {1 \over \epsilon}  - 2 \ln \left(-r^+-i 0^+ \right) \right) {1 \over r^+  + i 0^+ }, \nn
&=& {g^2 C_F \over 8 \pi^2}\left\{ {1 \over \epsilon}  \delta(r^+)   - 2 \left[ {\theta(r^+) \over r^+} \right]_{\dist}\right\}. \nn
\label{4.20}
\end{eqnarray}
The $1/\epsilon$ is an infrared divergence.

The heavy quark wavefunction graph vanishes on-shell when evaluated in pure dimensional regularization, because the integral is scaleless. One can verify that the graph has the structure $0=1/\epsilon_{\text{UV}}-1/\epsilon_{\text{IR}}$ with exactly the right coefficient to convert the infrared divergence in Eq.~(\ref{4.20}) into an ultraviolet divergence.

The wavefunction renormalization of the internal Wilson line is zero since $n^2=0$. 

The one-loop matrix element of the shape function operator is given by the sum of the tree graph, twice graph (a), graph (b) and the wavefunction graphs. From the infinite parts, we see that the renormalized operator $O_v$ is related to the bare operator by a convolution with the renormalization coefficient
\begin{eqnarray}
O_v^{(0)}\left(r^+ \right)
&=& \int_{-\infty}^\infty {{\rm d}\ell^+} \ Z\left (r^+ , \ell^+\right) O_v\left(\ell^+ \right) ,
\label{4.27}
\end{eqnarray}
where $O_v^{(0)}\left(r^+ \right)$ is the bare operator, $O_v\left(r^+\right)$ is the renormalized operator, and
\begin{eqnarray}
Z\left (r^+,\ell^+ \right)&=& \delta\left(r^+- \ell^+  \right) +{\alpha_s C_F \over 2 \pi \epsilon} \Biggl\{2
\left[{\theta(r^+ - \ell^+) \over  r^+ - \ell^+ }\right]_{\dist} \nn
&&+\left(1-  {1\over \epsilon}   \right)\delta\left(r^+-\ell^+ \right)  \Biggr\}.
\label{4.28}
\end{eqnarray}
The crucial difference from the renormalization of the deep inelastic structure function is that Eq.~(\ref{4.28}) does not have the restriction $l^+ > 0$, so the $\theta$-function is satisfied for $-\infty < \ell^+  < r^+ $, rather than $0 < \ell^+ < r^+ $.

Differentiating Eq.~(\ref{4.27}) with respect to $\mu$ gives the renormalization group equation for the shape function operator:
\begin{eqnarray}
\mu { {\rm d} \over {\rm d} \mu} O_v\left(r^+ \right) &=& - \int_{-\infty}^{\infty} {\rm d}\ell^+ \gamma\left(r^+,\ell^+\right)
O_v\left(\ell^+ \right) ,
\label{4.29}
\end{eqnarray}
where
\begin{eqnarray}
\gamma\left(r^+,\ell^+\right) &=&- {\alpha_s C_F \over  \pi } \Biggl\{
\delta\left(r^+-\ell^+ \right) + 2
\left[{\theta(r^+ - \ell^+) \over  r^+ - \ell^+ }\right]_{\dist} \Biggr\},\nn
\label{4.31}
\end{eqnarray}
using Eq.~(\ref{4.28}). The shape function $f(k^+,\mu)$ satisfies the same equation, since it is the matrix element of $O_v(k^+)$.

The convolution in the renormalization group equation Eq.~(\ref{4.29}) makes this equation difficult to solve and so far no solution exists in the literature. Methods that are used to solve a similar equation for the parton distribution functions do not work in this case because of the difference in the $\theta$-function mentioned above.

\subsection{Expressions for  $W_i^{(f)}$}\label{sec:final}

In this section we combine the results of the previous sections and give the final expressions for the scalar functions $W_i^{(f)}$, from which all differential decay rates can be obtained. As before, we define
\begin{eqnarray}
\lefteqn{W^{(f)}_i(\bn \cdot p, n \cdot p=p^+) = }\nn
&&\int \!{\rm d}k^+ C^{(f)}_i(\bn \cdot p;p^+-k^+,\mu) f(k^+,\mu)\,,
\label{final1}
\end{eqnarray}
where the shape function $f(k^+,\mu)$ is given by Eq.~(\ref{fdef}). The expressions for  $C^{(f)}_i(n \cdot p,p^+-k^+,\mu)$ at $\mu=\mu_2$ to first order in $\alpha_s$ can be obtained from Eqs.~(\ref{Wmatchu}), (\ref{Wmatchs}), (\ref{matchC}). As explained earlier, one can chose the matching scale from QCD onto SCET to be any scale $\mu_1 \sim m_b$ and in Section \ref{sect_matchcurrent} we gave results for both $\mu_1=m_b$ and $\mu_1 = \bn \cdot p$. In this sections we give all results with the choice $\mu_1 = m_b$. We find
\begin{eqnarray}
&&\label{final2}
\lefteqn{C_i^{(f)}(\bn \cdot p, w, \mu_2) =}\\
&&  S(\mu_2,m_b) \bigg[a_i^{(f)} {\cal C}(w,\mu_2)  + \frac{\alpha_s(m_b) C_F}{4 \pi} b_i^{(f)} \delta(w)  \bigg] + {\cal O}(\alpha_s^2)\nonumber\,,
\end{eqnarray}
where the coefficients $a_i^{(f)}$, $b_i^{(f)}$ are functions of $\bn \cdot p$ and $m_b$.
The function ${\cal C}(w,\mu_2)$ is universal and does not depend on the decay process. It is given in Eq.~(\ref{3.15}).

The scale factor $S(\mu_2,m_b)$ is the running of two SCET currents from the matching scale $\mu_1=m_b$ to the scale $\mu = \mu_2$ where the OPE is performed. We can  write
\begin{eqnarray}
S(\mu_2,m_b) = \exp 2 \left[\frac{r_0(z)}{\alpha_s(m_b)} + r_1(z) \right],\nonumber
\end{eqnarray}
where $r_0(z)$ and $r_1(z)$ are given in Eq.~(\ref{fdef2}) (one should pick $\mu_1=m_b$ in these expressions), and $z=\alpha_s(\mu_2)/\alpha_s(m_b)$. The remaining coefficients are different for the decay $B \to X_u \ell \bar \nu$ and $B \to X_s \gamma$. For $B \to X_u \ell \bar \nu$ we find 
\begin{eqnarray}
a_1^{(u)} &=& \frac{1}{4} ,\nn
a_2^{(u)} &=& \frac{1}{\bn \cdot \hat p}, \nn
a_3^{(u)} &=& \frac{1}{2 m_b \bn \cdot \hat p} , \nn
a_4^{(u)} &=& 0, \nn
a_5^{(u)} &=& -\frac{1}{2 m_b \bn \cdot \hat p}\,,\nonumber
\end{eqnarray}
where
\begin{eqnarray}
\bn \cdot \hat p = {\bn \cdot p \over m_b} = {m_b^2-q^2 \over m_b^2}.\nonumber
\end{eqnarray}
The coefficients $b_i^{(u)}$ are given by
\begin{eqnarray}
b_1^{(u)} &=&-\frac 1 2  g(\bn \cdot \hat p, m_b,m_b) - \log \bn \cdot \hat p\,\,\frac{3\bn \cdot \hat p-2}{2(1-\bn \cdot \hat p)}, \nn
b_2^{(u)} &=& -{2 g(\bn \cdot \hat p, m_b,m_b) \over \bn \cdot \hat p} +6{ \log \bn \cdot \hat p \over \bn \cdot \hat p}, \nn
b_3^{(u)} &=& -{g(\bn \cdot \hat p, m_b,m_b) \over m_b \bn \cdot \hat p} -{ \log \bn \cdot \hat p \over m_b \bn \cdot \hat p} \,\,\frac{3\bn \cdot \hat p-2}{1-\bn \cdot \hat p} ,\nn
b_4^{(u)} &=&- \frac{2}{m_b^2 \bn \cdot \hat p (1-\bn \cdot \hat p)} +2{ \log \bn \cdot \hat p \over m_b^2 \bn \cdot \hat p} \,\,\frac{1-2 \bn \cdot \hat p}{(1-\bn \cdot \hat p)^2},\nn
b_5^{(u)} &=& {g(\bn \cdot \hat p, m_b,m_b) \over m_b \bn \cdot \hat p}  + \frac{1}{m_b \bn \cdot \hat p (1-\bn \cdot \hat p)} 
\nn
&& \hspace{.3cm} - {\log  \bn \cdot \hat p \over m_b \bn \cdot \hat p} \,\,\frac{(3\bn \cdot \hat p -2)( \bn \cdot \hat p-2)}{(1-\bn \cdot \hat p)^2}.\nonumber
\end{eqnarray}
with 
\begin{eqnarray}
g(\bn \cdot p, m_b,m_b) = 2\ln^2(\bn \cdot \hat p)
  + 2 {\rm Li}_2(1\!-\!\bn \cdot \hat p)   + \frac{\pi^2}{12} + 6 .\nonumber
\end{eqnarray}

For the decay $B \to X_s \gamma$, $\bn \cdot p = m_b$. This simplifies the resulting expression significantly and we find 
\begin{eqnarray}
a_1^{(s)} &=& \frac{1}{4}, \nn
a_2^{(s)} &=& 0 ,\nn
a_3^{(s)} &=& \frac{1}{2 m_b} ,\nn
a_4^{(s)} &=&- \frac{3}{4 m_b^2} ,\nn
a_5^{(s)} &=& \frac{1}{2 m_b},\nn
a_\gamma^{(s)} &=& m_b.
\label{Cgammafinal1}
\end{eqnarray}
For the coefficients $b_i^{(s)}$:
\begin{eqnarray}
b_1^{(s)} &=&  -3 -  \frac{\pi^2}{24},\nn
b_2^{(s)} &=& 0 ,\nn
b_3^{(s)} &=&  -{6 \over m_b} -  \frac{\pi^2}{12m_b},\nn
b_4^{(s)} &=& \frac{10}{m_b^2}+\frac{\pi^2}{8 m_b^2},\nn
b_5^{(s)} &=&  -{6 \over m_b} -  \frac{\pi^2}{12m_b},\nn
b_\gamma^{(s)} &=& -\left(12+{\pi^2 \over 6}\right) m_b. 
\label{Cgammafinal2}
\end{eqnarray}

The above equations give $C_i^{(f)}(\bn \cdot p, w, \mu_2)$, and so must be integrated with the shape function $f(k^+,\mu_2)$ also renormalized at the scale $\mu_2$. One can instead convert to the shape function at some other scale (such as $\mu_3$) by using the renormalization group equation Eq.~(\ref{4.31}) for the shape function.

\section{$W^{(f)}_i$ in the phase space region $p^+ \gg \lqcd$.}
\label{sec:local}

As explained earlier, in regions of phase space where $p^+ \gg \lqcd$ an additional matching can be performed at the scale $\mu_3 = p^+$. Since the non-locality of the bilocal operator $O_v\left(r^+ \right)$ was set by the scale $p^+$, the operators below the scale $\mu_3$ will be local operators. In this section we will match onto these operators at one loop and then determine their running.

\subsection{Matching onto local operators}

The bilocal operator $O_v(p^+)$ below $\mu_3$ can be expanded in terms of local operators $O_m$,
\begin{eqnarray}
O_v(p^+) = \sum_m{\cal D}_m(p^+,\mu) O_m,
\label{16.01}
\end{eqnarray}
with coefficients ${\cal D}_m(p^+,\mu)$,
where (at tree level)
\begin{eqnarray}
O_m=  \bar b_v (in \cdot D)^m b_v \,.\nonumber
\end{eqnarray}
The matrix elements of $O_m$ are of order $\lqcd^m$. The coefficients ${\cal D}_m$ of are of order $(p^+)^{-m-1}$. The subleading terms in the endpoint region have coefficients suppressed by powers of $p^+/m_b$.

The local operators $O_m$ are related to the moments of $O_v(p^+)$.
The matrix elements of the first few local operators are given by
\begin{eqnarray}
\langle B | \bar b_v b_v | B \rangle &=& 2m_B\,,\nn
\langle B | \bar b_v( i n \cdot D) b_v | B \rangle &=& 0\,,\nn
\langle B | \bar b_v (i n \cdot D)^2 b_v | B \rangle &=& - \frac{2m_B}{3} \,\,\lambda_1(\mu) \,,\nn
\langle B | \bar b_v (i n \cdot D)^3 b_v | B \rangle &=& - \frac{2m_B}{3} \,\,\rho_1(\mu),\nonumber
\end{eqnarray}
to leading order in $1/m_b$. $\lambda_1$ and $\rho_1$ are given by the matrix elements of the kinetic and Darwin terms
\begin{eqnarray}
\langle B | \bar b_v D^2 b_v | B \rangle &=&-4 m_B \lambda_1(\mu) ,\nn
g v_\beta \langle B | \bar b_v \left( D_\alpha G^{\alpha \beta} \right)  b_v | B \rangle &=& - 4 m_B\rho_1(\mu).\nonumber
\end{eqnarray}

Since the scale $\mu_3$ is now a large scale, the matrix element of the operator $O_v(r^+)$ can be calculated perturbatively, using on-shell partonic states with residual momentum $k=0$. At tree level the matching Eq.~(\ref{16.01}) is trivial since $O_v(p^+) = \bar b_v \delta(i n \cdot D+ p^+) b_v  + {\cal O}(\alpha_s)$, and we find
\begin{eqnarray}
{\cal D}_m(p^+) &=& \frac{1}{m!} \left(\frac{{\rm d}}{{\rm d}p^+} \right)^m \delta(p^+) + {\cal O}(\alpha_s)\,.\nonumber
\end{eqnarray}

The order $\alpha_s$ matrix element is given by the finite parts of Eq.~(\ref{4.03}), twice Eq.~(\ref{4.14}) and Eq.~(\ref{4.20}),
\begin{eqnarray}
&&\me{b_v}{O_v(p^+)}{b_v}= \delta(p^+) 
- {\alpha_s(\mu) C_F \over 4 \pi} \Biggl[ {\pi^2 \over 6}   \delta(p^+)
\nn
&&
+ 4 \left[ {\theta(p^+)\over p^+}\right]_{\dist} + 8 \left[ { \ln (p^+/\mu)\ \theta(p^+)\over p^+}\right]_{\dist} \Biggr]\,.
\label{4.37}
\end{eqnarray}
Since $k^+=0$, the matrix elements of $O_m$ vanish except for $m=0$, so we get
\begin{eqnarray}
&&
{\cal D}_0(p^+,\mu) =  \delta(p^+) 
- {\alpha_s C_F \over 4 \pi} \Biggl[ {\pi^2 \over 6}   \delta(p^+)
\nn
&&
+ 4 \left[ {\theta(p^+)\over p^+}\right]_{\dist} + 8 \left[ { \ln (p^+/\mu)\ \theta(p^+)\over p^+}\right]_{\dist} \Biggr]\,.
\label{4.37a}
\end{eqnarray}
This matrix element gives the parton level value for the shape function,
\begin{eqnarray}
f_{\text{part}}(k^+,\mu)&=&f_{\text{part}}^{(0)}(k^+,\mu) + {\alpha_s(\mu) C_F \over 4 \pi} f_{\text{part}}^{(1)}(k^+,\mu),\nn
f_{\text{part}}^{(0)}(k^+,\mu) &=& \delta(k^+),\nn 
f_{\text{part}}^{(1)}(k^+,\mu) &=&- {\pi^2 \over 6}   \delta(k^+)
 - 4 \left[ {\theta(k^+)\over k^+}\right]_{\dist} \nn
 && - 8 \left[ { \ln (k^+/\mu)\ \theta(k^+)\over k^+}\right]_{\dist} .
\label{4.37b}
\end{eqnarray}

To determine the coefficients ${\cal D}_m$ for $m>0$ requires evaluating the matrix element Eq.~(\ref{4.37}) for non-zero $k^+$, with $2k\cdot v = k^++k^-=0$, so that the quark is still on-shell. It can be determined, following the arguments in Ref~\cite{MW}, in terms of the lowest order coefficient. The matrix element Eq.~(\ref{4.37}) for non-zero $k^+$ is given by the replacement $p^+ \to p^+ + k^+$, which follows from reparametrization invariance~\cite{RPI}, so that
\begin{eqnarray}
&&\me{b_v(k^+)}{O_v(p^+)}{b_v(k^+)} = {\cal D}_0(p^++k^+,\mu)\nn
&=& \sum_m  \frac{(k^+)^m}{m!} \left(\frac{{\rm d}}{{\rm d}p^+} \right)^m{\cal D}_0(p^+,\mu).\nonumber
\end{eqnarray}
The matrix elements of $\bar b_v (in \cdot D)^m b_v$ are their tree level value $(k^+)^n$, since loop graphs are scaleless and vanish, so
\begin{eqnarray}
{\cal D}_m(p^+,\mu)=\frac{1}{m!} \left(\frac{{\rm d}}{{\rm d}p^+} \right)^m{\cal D}_0(p^+,\mu).
\label{simple}
\end{eqnarray}
The matching condition Eq.~(\ref{simple}) is evaluated at a scale $\mu \sim p^+$ to minimize logarithms.  At one loop, one can also generate four-quark operators; the matching coefficients for these operators is not computed here.

\subsubsection{Local Operators in Deep Inelastic Scattering}

In deep inelastic scattering, the on-shell matrix element of the parton distribution operator vanishes, so the analog of Eq.~(\ref{4.37b}) is $f^{(1)}=0$. An on-shell external state has $k^+ \not =0$, $k^-=0$. The momentum $p^+$ in the operator and the momentum $k^+$ of the external state enter the loop integral over the $+$ component of the loop integral, but the $-$ component of the loop integral is scaleless, and vanishes. In the $B$ meson shape function matrix element Eq.~(\ref{4.37}), the heavy quark propagator $v \cdot \ell $ mixes the $+$ and $-$ components of the loop momentum $\ell$, so the scale $p^+$ enters both loop integrals, and the $\ell^-$ integral is no longer scaleless.

\subsection{Running of the local operators}

Since the matrix elements of the local $\bar b_v (in \cdot D)^m b_v$ operators are equal to the usual HQET operator matrix elements that define $\lambda_1$, $\rho_1$, etc.,
the renormalization group equations for the matrix elements of these two sets of operators must be identical. One can check that the running of the local $\bar b_v (i n \cdot D)^m b_v$ operators reproduces the known renormalization group evolution for $\lambda_1$ and $\rho_1$~\cite{localrunning2} for $m=2,3$, respectively. There is an important difference here between the evolution of the local $\bar b_v(i n \cdot D)^m b_v$ operators and the corresponding evolution of local operators in deep inelastic scattering. In deep inelastic scattering, the evolution of the bilocal parton distribution operator is equivalent to the evolution of the local twist-two operators; the twist-two anomalous dimensions are given by the moments of the Altarelli-Parisi evolution kernel for the parton distribution. This is not the case for the shape function operator and its moments. The shape function anomalous dimension at one-loop depends only on $C_F$, whereas the anomalous dimension of $\rho_1$, which is the same as the anomalous dimension of the Darwin term, depends on $C_A$~\cite{localrunning2}.

\begin{figure}
\def\size{3.5 cm}
\hbox{\hspace{1cm}\vbox{\hbox to \size {\hfil \includegraphics[width=3cm]{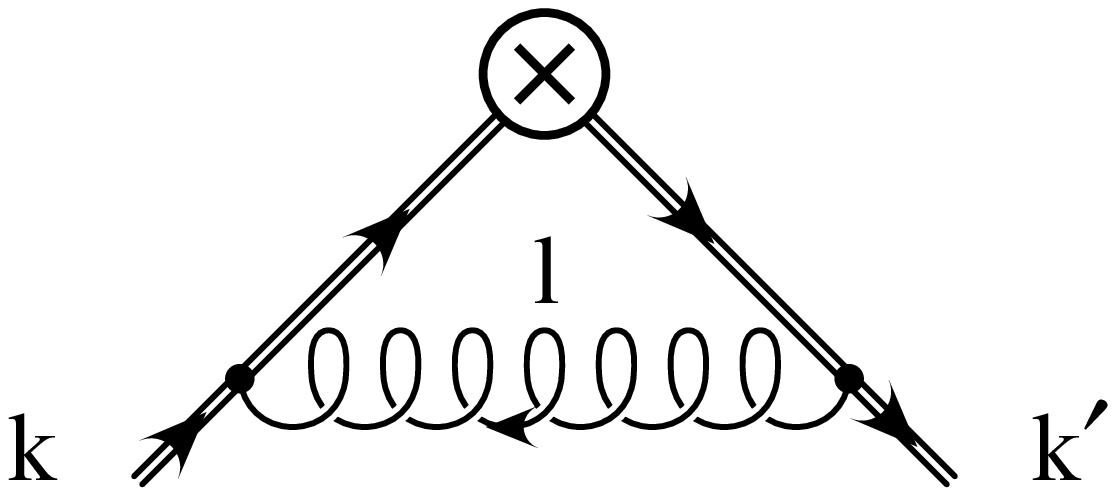} \hfil }\hbox to \size {\hfil(a)\hfil}}
\vbox{\hbox to \size {\hfil \includegraphics[width=3cm]{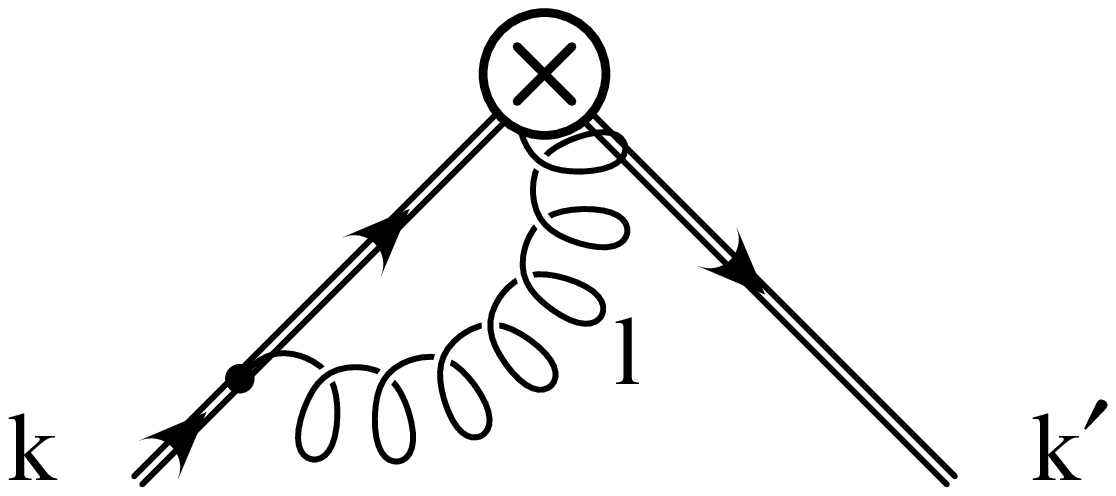} \hfil}\hbox to \size {\hfil(b)\hfil}}
}
\caption{One loop corrections to the local operator. \label{fig5}}
\end{figure}

The two diagrams shown in Fig.~\ref{fig5} contribute to the renormalization of the local operators $O_m$.
For the first diagram we find for the forward matrix element ($k^\prime=k$)
\begin{eqnarray}
I_a
&=&- i g^2 C_F\int {{\rm d}^d \ell \over \left(2 \pi \right)^d} {\left[ n \cdot(k+\ell) \right]^m \over \left[ v \cdot \ell \right]^2 \ell^2}= -{g^2C_F \over 8 \pi^2 \epsilon} \left( n\cdot k\right)^m,\nn
\label{5.04}
\end{eqnarray}
where we have only kept the divergent part of the graph and used the equation of motion $k \cdot v=0$ for the external states since we are only working to lowest order in $1/m_b$.  The graph Fig.~\ref{fig5}(b) gives
\begin{eqnarray}
I_b &=& ig^2 C_F \sum_{j=1}^m
\int {{\rm d}^d \ell \over \left(2 \pi \right)^d} 
\left( n \cdot k \right)^{j-1} \frac{\left[ n \cdot (k+\ell) \right]^{m-j}}{v \cdot \ell \,\,\ell^2} 
=  0,\nonumber
\end{eqnarray}
since all the divergent terms are proportional to powers of  $(k\cdot v)$, and vanish.
The wavefunction graph gives
\begin{eqnarray}
- i {g^2 \over 8 \pi^2 \epsilon}  (k \cdot v) = i (k \cdot v) \delta Z,\nonumber
\end{eqnarray}
so that the wavefunction graph cancels Eq.~(\ref{5.04}), when multiplied by the tree-level matrix element 
\begin{eqnarray}
\me{k}{O_m}{k} &=& \left(n \cdot k \right)^m\,,\nonumber
\end{eqnarray}
$I_a - \delta Z (n \cdot k)^m=0$. The net result is that
the operator $\bar b_v (in \cdot D)^m b_v$ has no anomalous dimension from the graphs in Fig.~\ref{fig5} at leading order in $1/m_b$.

\begin{figure}
\includegraphics[width=3cm]{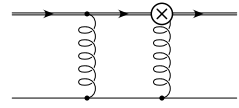}
\caption{One loop running of the local operator. \label{fig:box}}
\end{figure}

While this gives the correct result for the running of the local operators $O_m$ for $m<3$, it does not reproduce the known result for $m\ge3$. For the latter case, however, there are additional graphs such as those in Fig.~(\ref{fig:box}), which are divergent for $m\ge 3$ and renormalize the local operator. 
These box graphs connect the heavy quark to a light quark, and mix into four fermion operators. Such graphs are absent for deep inelastic scattering, where there is a twist expansion. Fermion fields have twist one, and two quark operators with twist two cannot mix with four-quark operators which have twist four, so the box graph Fig.~\ref{fig:box} does not contribute to the anomalous dimension. In $B$ decay, there is no analogous twist expansion, and the mixing graphs of Fig.~(\ref{fig:box}) are present.\footnote{The Feynman integrals for Fig.~\ref{fig:box} for deep inelastic scattering involve $\xslash n \xslash n =0$, whereas for $B$ decay involve $\xslash n \xslash v \not=0$.} They are precisely the graphs computed in Ref.~\cite{localrunning2} which give the anomalous dimension of the Darwin term whose matrix element is $\rho_1$, and allow the dimension six operator $\bar b_v (in \cdot D)^3 b_v$ to mix into the dimension six operator $\bar b_v  b_v \bar q q$.

\subsection{Expressions for $W_i^{(f)}$ }
\label{sec:finalW2}

The final expressions for $W_i^{(f)}$ in the region $\Delta \gg \lqcd$ are given by combining Eq.~(\ref{final1}) with the expansion of the shape function operator, Eq.~(\ref{16.01}). They are given by
\begin{eqnarray}
&&W_i(\bn \cdot p,n \cdot p =p^+) 
= S(\mu_2,m_b) 
\\
&&
\hspace{.5cm}
\times
a_i^{(f)} \Biggl\{\delta(p^+) 
+ \frac{\alpha_s C_F}{4 \pi} \Biggl[-4 \left[ \frac{\ln(p^+/\mu_2)\theta(p^+)}{p^+}\right]_{\mu_2} 
\nn
&&
\hspace{.5cm}
+ \left(4 \ln \frac{\bn \cdot p}{\mu_2} - 7 \right) \left[\frac{\theta(p^+)}{p^+} \right]_{\mu_2} 
+ \left(2 \ln^2 \frac{\bn \cdot p}{\mu_2} 
\right.\nn
&&
\left.
\hspace{1.5cm}
- 3 \ln \frac{\bn \cdot p}{\mu_2}  
+ 7 - \frac{7 \pi^2}{6} + \frac{b_i^{(f)}}{a_i^{(f)}} \right) \delta(p^+) \Biggr] \Bigg\}\nonumber
\end{eqnarray}
where $S(\mu_2,m_b)$, $a_i^{(f)}$ and $b_i^{(f)}$ are the same as in section \ref{sec:final}.

\section{Constraints on the anomalous dimensions}

There is a relation between the anomalous dimensions in the effective theory similar to the one obtained for deep inelastic scattering~\cite{dis}. It is simplest to write the relations in terms of moments. The theory below $\mu_1$ has SCET currents, and depends on the label momentum $\bn \cdot p$. Anomalous dimensions can depend on $\bn \cdot p$, but not on any infrared scales in this theory such as $n \cdot p$, and so cannot depend on the moment $N$. In the theory below $\mu_2$,  the effective theory is written in terms of shape function operators that depend on the scale $p^+=m_b(1-x)=m_b/\bN$. The scale $p^-$ has been integrated out, so the anomalous dimension is only a function of $m_b/\bN$.
Finally, in the theory below $\mu_3$, the scale $p^+$ has also been integrated out, and the theory only knows about $\bN$, since the local operator depends on the moment. The anomalous dimensions $\gamma_i$ in the three theories below $\mu_i$ can be written as
\begin{eqnarray}
2\gamma_1 (\mu) &=& f\left({\mu \over \bn \cdot p}, \alpha_s(\mu) \right), \nn[5pt]
\gamma_2 (\mu) &=& f\left({ \bN \mu \over m_b}, \alpha_s(\mu) \right), \nn[5pt]
\gamma_3 (\mu) &=& f\left(\bN , \alpha_s(\mu) \right). \nonumber
\end{eqnarray}
We have used twice the anomalous dimension $\gamma_1$ of the SCET current below $\mu_1$, since the time-ordered product involves two currents.

The matching conditions at $\mu_2$ depend only on the invariant mass $p^2$ of the hadronic final state which is integrated out, so the matching coefficients are functions $C_2(p^2/\mu^2)=C_2(m_b \bn \cdot p/(\bN \mu^2))$. The matching conditions at $\mu_3$ depend only on the scale $p^+$ which is integrated out, so the matching coefficients are functions $C_3(p^+/\mu)=C_3(m_b/(\bN \mu))$. Finally, using the relation that the derivative of the matching coefficients with respect to $\mu$  is equal to the difference of the anomalous dimensions on either side, one finds that the anomalous dimensions: (1) must be linear in $\ln \mu$, and (2) must have the form
\begin{eqnarray}
2 \gamma_1 (\mu) &=&A\left( \alpha_s(\mu) \right)\ln {\mu \over \bn \cdot p}  +B_1\left( \alpha_s(\mu) \right), \nn[5pt]
\gamma_2 (\mu) &=&-A\left( \alpha_s(\mu) \right)\ln { \bN \mu \over m_b}  +B_2\left( \alpha_s(\mu) \right),\nn[5pt]
\gamma_3 (\mu) &=&0\ln { \bN}  +B_3\left( \alpha_s(\mu) \right),
\label{20.35}
\end{eqnarray}
which restricts the coefficients of the $\ln \mu$ terms. The results Eqs.~(\ref{adim2},\ref{4.38}) satisfy these constraints, with $A=-2 \alpha_s C_F/\pi$. The third relation shows that the anomalous dimension of $\bar b_v(in \cdot D)^m b_v$ should not have a $\ln m$ term as $m \to \infty$. 

The anomalous dimensions $\gamma_{1,2}$ are used for $\mu_1 \ge \mu \ge \mu_2$ and $\mu_2 \ge \mu \ge \mu_3$, respectively. The $\ln \mu$ term in $\gamma_1$ is not a large logarithm at the upper endpoint $\mu_1$, and the $\ln \mu$ term in $\gamma_3$ is not a large logarithm at the lower endpoint $\mu_3$. At these limits, the effective theories match with
full QCD above $\mu_1$, which has no $\ln \mu$ anomalous dimension, and the theory with local operators below $\mu_3$, which has no $\ln \bar N$ anomalous dimension.

\subsection{Relation to the factorization form of the anomalous dimension}

Korchemsky and Sterman~\cite{KorSter} have derived an alternative form for the running  of the $B \to X_s \gamma$ moments between the scales $m_b$ and $m_b/\bN$. They have a renormalization group evolution of the form
\begin{eqnarray}
C\left({m_b \over \bN}\right) &=& C(m_b) e^\Lambda ,\nonumber
\end{eqnarray}
where
\begin{eqnarray}
\Lambda =\int_{1/\bN}^1 {{\rm d} y \over y} \Biggl[  \Gamma(\alpha(m_b y) )+ \gamma(\alpha(m_b \sqrt y)) \nn
+ \int_{m_b y}^{m \sqrt y}  {{\rm d} \mu \over \mu}
2 \Gamma_c(\alpha(\mu))   \Biggr] .
\label{20.30}
\end{eqnarray}
Changing the order of integration gives
\begin{eqnarray}
\Lambda &=&  \int_{m /\sqrt{ \bN }}^{m_b}  {{\rm d} \mu \over \mu} \left[ \ln {m_b \over \mu} 2 \Gamma_c(\alpha(\mu)) + \Gamma(\alpha(\mu)) + 2 \gamma(\alpha(\mu))\right]\nn 
&&
+  \int_{m_b /\bN}^{m_b/\sqrt {\bN}}  {{\rm d} \mu \over \mu} \left[
  \ln {\mu \bN \over m_b}  2 \Gamma_c(\alpha(\mu))+  \Gamma(\alpha(\mu))\right],\nonumber
\end{eqnarray}
and so is equivalent to integrating the anomalous dimension $2\gamma_1$ between $m_b$ and $m_b/\sqrt {\bN} $ and the anomalous dimensions $\gamma_2$  between $m_b/\sqrt {\bN}$ and $m_b/\bN$ where
\begin{eqnarray}
2 \gamma_1 &=& -2 \ln {\mu \over m_b} \Gamma_c(\alpha(\mu)) + \Gamma(\alpha(\mu)) + 2 \gamma(\alpha(\mu)),\nn
\gamma_2 &=& 2 \ln { \bN \mu \over m_b }  \Gamma_c(\alpha(\mu))+  \Gamma(\alpha(\mu)) , \nonumber
\end{eqnarray}
and so satisfies the constraint on the coefficients of the $\ln \mu$ terms in Eq.~(\ref{20.35}), with $\Gamma_c=\alpha_s C_F/\pi$. Alternatively, the integrations from $\mu_1 \to \mu_2$ and $\mu_2 \to \mu_3$ can be combined into the single integral Eq.~(\ref{20.30}) since the constraint Eq.~(\ref{20.35}) is satisfied.

The double integral form Eq.~(\ref{20.30}) has a Landau pole singularity for large moments. As for deep inelastic scattering, this Landau pole singularity is resolved by the effective theory computation~\cite{dis}. For large enough moments, the scales $\mu_2=m_b/\sqrt{\bN}$ and  $\mu_3=m_b/\bN$ get smaller than $\mu_4=\lqcd$, and one needs to stop the renormalization group evolution at $\mu_4$. For $B$ decays the condition $\mu_3 > \mu_4$, which is required for Eq.~(\ref{20.30}) to be valid is only true for the first two moments.

\section{Applications}
\label{sec:app}

\subsection{Moments of the shape function}

It is generally believed that moments of the structure function $f(k_+)$ are related to local operators in HQET. This relation is used to construct models of the structure function using input from matrix elements of local operators extracted from inclusive $B \to X_c \ell \bar \nu$ decays or to constrain fits to the shape function measured from data. We have already seen that this relation cannot be correct, since the shape function and local operators have different anomalous dimensions. In this section we show in more detail why this relation is incorrect.

Moments of the bare operator $O^{(0)}_v(k_+)$ are given by bare local operators,
\begin{eqnarray}
M_n^{(0)} &=& \int_{-\infty}^\infty {\rm d}k_+\left(-k_+\right)^n O^{(0)}_v(k_+) \nn
&=& \bar b_v (i n \cdot D)^n b_v, \nonumber
\end{eqnarray}
but the corresponding relation {\it does not} hold for renormalized operators. The infinite moments, over $-\infty \le r^+ \le \infty$ of the renormalized operator $O_v(p^+)$  are singular. One way to see this is to take the infinite moments of the renormalization group evolution Eq.~(\ref{4.29}).  The integral over $\gamma(r^+,\ell^+)$ does not converge at $r^+=\infty$, so the infinite moments require additional regularization.

To gain more insight into the running of these moments without encountering the divergence, one can also study the evolution of half-infinite moments, defined by
\begin{eqnarray}
M_N^+ \left( O \right) &=& \int_{0}^\infty {\rm d} r^+ \left(r^+\right)^{N-1}\ O\left(r^+ \right),\nn
M_N^- \left( O \right) &=& \int_{-\infty}^0 {\rm d} r^+ \left(r^+\right)^{N-1}\ O\left(r^+ \right)\,. \nonumber
\end{eqnarray}
Consider the half infinite moment $M_N^- \left( O \right)$. We find
\begin{eqnarray}
&&\hspace{-.8cm}\mu { {\rm d} \over {\rm d} \mu} M_N^- \left( O \right)\nn
&=&-\int_{-\infty}^0 {\rm d} r^+ \left(r^+\right)^{N-1} \int_{-\infty}^\infty
 {\rm d}\ell^+ \gamma\left ( r^+ , \ell^+ \right) O\left(\ell^+ \right) \nn
&=& -\int_{-\infty}^\infty  {{\rm d}\ell^+} O\left(\ell^+ \right)
 \int_{-\infty}^0 {\rm d} r^+ \left(r^+\right)^{N-1}\ \gamma\left ( r^+ , \ell^+ \right)\,.\nonumber
\end{eqnarray}
Since $\gamma(r^+,\ell^+)=0$ unless $\ell^+ \le r^+$, and $r^+ \le 0$ in the integration region, we can restrict the $\ell^+$ integration to $-\infty \le \ell^+ \le0$. Let $r^+=z\ell^+$:
\begin{eqnarray}
&&\hspace{-.8cm}\mu { {\rm d} \over {\rm d} \mu} M_N^- \left( O \right) \nn
&=&
{\alpha_s C_F \over \pi} \int_{-\infty}^0  {\rm d}\ell^+ \left( \ell^+ \right)^{N-1} \ O\left(\ell^+ \right) \nn
&& \hspace{.5cm}
\times  \int_0^{\infty} {\rm d}z z ^{N-1}\  \bigg\{
\delta\left(1-z \right)  
\nn
&& \hspace{.5cm}
+2
\left[{\theta\left(z\ell^+- \ell^+ > \xi \right) \over 1-z  } +  \delta(1-z) \ln {\xi\over \mu}\right] \bigg\} \nn
 &=&
{\alpha_s C_F \over \pi} \int_{-\infty}^0  {\rm d}\ell^+ \left( \ell^+ \right)^{N-1} \ O\left(\ell^+ \right)  
\nn
&& \hspace{.5cm}
\times
\left[
1 + 2 \ln {-\ell^+ \over \mu} -  2 \sum_{j=1}^{N-1}{1 \over j}  \right].
\label{MN-}
\end{eqnarray}
This shows that the half-infinite moments $M_N^- \left( O \right) $ are not multiplicatively renormalized, because of the $\ln (-\ell^+/\mu)$ term in Eq.~(\ref{MN-}). 
The half-infinite moment $M_N^+ \left( O \right)$ has the divergence at $r^+ \to  \infty$ noted earlier. For moments of the shape function only the half infinite moment $M_N^+\left(O\right)$ is relevant, since $f(r^+)$ vanishes for $r^+ <  -\bar \Lambda$. Thus, moments of the shape function are not defined without an additional renormalization prescription and are not directly related to matrix elements of local operators once radiative corrections are included. Models for the shape function should therefore not be constrained to have width $\sigma^2 = -\lambda_1/(3m_b^2)$ for example. The parameters of a given model have to be kept arbitrary and determined directly from a fit to the data.

Since the shape function is not equivalent to local operators at order $\alpha_s$, the moments of the shape function are not given by the matrix elements of the local operators. Instead, one has to compute the matching correction onto the local operators, as in Eq.~(\ref{16.01}). In particular the first moment of the shape function (because of the definition of the moments with powers of $N-1$, the first moment is the normalization of the shape function) is not unity. The matrix element of $O_0=\bar b_v b_v$ is unity to all orders in $\alpha_s$ and leading order in $1/m_b$, by heavy quark symmetry, so the first moment of the shape function is given by the first moment of ${\cal D}_0$, the matching coefficient onto the lowest order operator $O_0$. ${\cal D}_0$ has a non-zero first moment, as can be seen from Eq.~(\ref{17.21}).

Note that low moments of the photon energy spectrum in $B \to X_s \gamma$ are still given by the well known matrix elements of local operators $\lambda_1$, $\rho_1$ etc. This is because for these moments one integrates over the entire region of phase space and the traditional OPE is therefore applicable. High moments of the photon spectrum no longer match onto local operators, and are discussed in Sec.~\ref{sec:rgmom}.

\subsection{Expressions for differential decay rates}\label{sec:wtoalpha}

Using the results of this paper we can obtain expressions for differential decay rates to order $\alpha_s$. In this section we will give results for the photon energy spectrum $d\Gamma/dx_\gamma$ in the limit $x_\gamma \to 1$ for the decay $B \to X_s \gamma$  and the double differential decay rate $d\Gamma/(dz dx_\ell)$ in the limit $x_\ell \to 1$ for the decay $B \to X_u \ell \bar \nu$. For both decays we give the results in the region $m_b \gg m_b(1-x_{\ell,\gamma}) \gg \lqcd$ where an expansion in local operators is valid, and in the shape function region $m_b(1-x_{\ell,\gamma}) \sim \lqcd$.

\subsubsection{$B \to X_s \gamma$ for $m_b \gg m_b(1-x_{\gamma}) \gg \lqcd$}
\label{sec:diffs1}

In this region of phase space, the differential decay rate can be expanded around the limit $x_\gamma = 1$, but we are not in the shape function region and the non-perturbative effects are still obtained by matrix elements of local operators. The photon energy spectrum is given by
\begin{eqnarray}
\frac{1}{\Gamma_0^s}\frac{{\rm d}\Gamma^s}{{\rm d} x_\gamma}(x_\gamma) = W_\gamma\left(\bn \cdot p, p^+=m_b(1-x_\gamma)\right)
\end{eqnarray}
where $W_\gamma$ was defined in Eq.~(\ref{17.20}). Using the expression of $W$ as a convolution of a  Wilson coefficient $C_\gamma$ with the bilocal operator $O_v$, (\ref{final1}) together with the expansion of the bilocal operator in terms of local operators given in (\ref{16.01}) we find
\begin{eqnarray}
W_\gamma &=& \int dr^+ C_\gamma(m_b(1-x_\gamma) - r^+,\mu_2) 
\nn
&& \qquad 
\times \sum_m {\cal D}_m(r^+,\mu_2) {\me{\bar B}{O_m}{\bar B} \over 2 m_B}.
\label{wgamma}
\end{eqnarray}
In writing this expression we have ignored the running between the scales $\mu_2$ and $\mu_3$ and are therefore performing the matching onto the bilocal operator and the matching onto the local operator at the same scale. This allows one to give an expression for the decay spectrum ${\rm d}\Gamma/{\rm d}x_\gamma$ directly, rather than for its moments. The complete expression for the moments is given in Section~\ref{sec:rgmom}.

Keeping only the lowest  dimensional operator $\me{B}{O_0}{B} = 2m_B$ and using the explicit expressions for the Wilson coefficients $C_\gamma$ and ${\cal D}_0$ given in Eqs.~(\ref{final2}), (\ref{Cgammafinal1}), (\ref{Cgammafinal2}) and (\ref{4.37a}) we find
\begin{eqnarray}
\frac{1}{\Gamma_0^s}\frac{{\rm d}\Gamma}{{\rm d}x_\gamma} &=& S(\mu_2,m_b) \Bigg\{\delta(1-x_\gamma) 
\label{17.16}
\\
&& 
- \frac{\alpha_s C_F}{4\pi} \Biggl[ 4 \left({ \ln(1-x_\gamma) \over 1-x_\gamma } \right)_+ 
+
7 \left({ 1 \over 1-x_\gamma } \right)_+ \nn
&&\hspace{-1cm} - \left( \ln^2{\mu_2^2 \over m_b^2} + 5\ln{\mu_2^2 \over m_b^2} - 5 - {4 \pi^2 \over 3} \right) \delta(1-x_\gamma) \Biggr] \Bigg\}+\ldots,\nonumber
\end{eqnarray}
where $\ldots$ denotes perturbative terms of order $\alpha_s^2$ as well as terms of order $\alpha_s$ which are not singular as $x_\gamma \to 1$. Furthermore, we have not given explicitly the power suppressed terms proportional to matrix elements of higher dimensional operators, which can be included by retaining the operators $O_{m>1}$. The explicit logarithms of $\mu_2/m_b$ in Eq.~(\ref{17.16}) are canceled by the $\mu_2$ dependence of $S(\mu_2,m_b)$. To fixed order in $\alpha_s$ our results agree with the known perturbative corrections (see the next subsection), but we improve this result by including the Sudakov logarithms originating from running between the scales $m_b$ and $\mu_2 \sim 1.5$ GeV.

\subsubsection{The $B \to X_s \gamma$ spectrum to order $\alpha_s$ ignoring renormalization group evolution}

The photon spectrum in $B \to X_s \gamma$ decay to order $\alpha_s$ ignoring renormalization group evolution is given by using Eq.~(\ref{17.16}), and expanding $S(\mu_2,m_b)$ to order $\alpha_s$:
\begin{eqnarray}
\frac{1}{\Gamma_0^s}\frac{{\rm d}\Gamma}{{\rm d}x_\gamma} &=& \delta(1-x_\gamma) - \frac{\alpha_s C_F}{4\pi} \Biggl\{ 4 \left[{ \ln(1-x_\gamma) \over 1-x_\gamma } \right]_+ \nn
&& +
7 \left[{ 1 \over 1-x_\gamma } \right]_+ 
+ \left[ 5 + {4 \pi^2 \over 3} \right] \delta(1-x_\gamma) \Biggr\}
\nn
&&\hspace{-1cm} 
+{\alpha_s C_F \over 4 \pi}E(x_\gamma),\nn
\label{17.16a}
\end{eqnarray}
up to terms $E(x_\gamma)$ which have vanishing moments as $N \to \infty$. The scale at which $\alpha_s$ is evaluated is not determined by this fixed order result.

A comparison of Eq.~(\ref{17.16a}) with the complete order $\alpha_s$ expression~\cite{BFL} shows that the terms we have computed agree, and the remaining terms are
\begin{eqnarray}
E(x) &=& 7+ x - 2x^2-2(1+x) \ln(1-x), \nonumber
\end{eqnarray}
with moment
\begin{eqnarray}
M_N\left[E(x)\right] &=& {4 N +2 \over N(N+1)} H_N +{7 \over N} - {N^2-N-4 \over (N+1)^2(N+2)},\nonumber
\end{eqnarray}
which vanishes as $N \to \infty$. Here $H_N$ is the harmonic number
\begin{eqnarray*}
H_N &=& \sum_{j=1}^N {1 \over j}.
\end{eqnarray*}

\subsubsection{$B \to X_s \gamma$ for $m_b(1-x_{\gamma}) \sim \lqcd$}

In the shape function region, the expansion of the bilocal operator in terms of local operators can not be performed, and we are left with
\begin{eqnarray}
W_\gamma &=& \int dr^+ C_\gamma(p^+ - k^+,\mu_2) f(k^+,\mu_2)
\end{eqnarray}
where $p^+ = m_b(1-x_\gamma)$. 
Using the expression for $C_\gamma$ given in Eq.~(\ref{final2}) we find
\begin{eqnarray}
\frac{1}{\Gamma_0^s}\frac{{\rm d}\Gamma^s}{{\rm } x_\gamma} &=& m_b \,\,S(\mu_2,m_b) \Bigg[
f(p^+,\mu_2) 
\\
&&
+ \frac{\alpha_s C_F}{4 \pi}  \int \!\! dk^+ G_s(p^+-k^+,\mu_2) f(k^+,\mu_2)\Bigg]\,,\nonumber
\end{eqnarray}
where 
\begin{eqnarray}
G_s(\omega,\mu) &=&  \left( \frac{1}{2} \ln^2 \frac{\mu^2}{m_b^2}+   \frac{3}{2} \ln\frac{\mu^2}{m_b^2} - 5 - \frac{7\pi^2}{6} \right)\delta(\omega)
\nn
&&
+ 4 \left( 
\frac{\ln \frac{\omega}{\mu}}{\omega}\right)_{\mu} - \left(2 \ln \frac{\mu^2}{m_b^2} + 3 \right) \left( \frac{1}{\omega}\right)_\mu .\nn
\end{eqnarray}

Note that this expression does not agree with previous prescriptions to obtain the photon energy spectrum in the shape function region \cite{dFN}, where the perturbative spectrum is evaluated with a shifted $b$ quark mass $m_b \to m_b + k^+$ and $k^+$ is then convoluted with the shape function $f(k^+)$. The reason for this disagreement is discussed in more detail in Section \ref{sec:diffconv}.

\subsubsection{$B \to X_u \ell \bar \nu$ for $m_b \gg m_b(1-x_{\ell}) \gg \lqcd$}

Any differential decay rates can be obtained from the general expression of the triple differential rate given in Eq.~(\ref{diffrate}). As an illustration, we present results for the double differential decay rate ${\rm d}\Gamma^u/({\rm d}z {\rm d}x_\ell)$ in the limit $x_\ell \to 1$. Using $\hat p^2  \ll 1$, we find for this double differential rate
\begin{eqnarray}
\frac{1}{\Gamma_0^u} \frac{{\rm d}\Gamma^u}{{\rm d}z {\rm d}x_\ell} = 12 z (1-z) \! \int_0^{m_b(1-x_\ell)} \!\!\!\!\!\!\!\!\!\!\!dp^+ \left[ 2W^u_1 + m_b z W^u_3
\right].\nn
\end{eqnarray}
As for the decay $B \to X_s \gamma$, one can use the expression for $W_i^{(u)}$ as a convolution of a  Wilson coefficient $C_i^{(u)}$ with the bilocal operator $O_v$, Eq.~(\ref{final1}), together with the expansion of the bilocal operator in terms of local operators given in Eq.~(\ref{16.01}). Using the results given in Section \ref{sec:final} for the Wilson coefficients $C_i^u$ together with Appendix \ref{app:dist} to perform the integration over the $\mu$ distributions in the $C_i^u$ we find
\begin{eqnarray}
\frac{1}{\Gamma_0^u} \frac{{\rm d}\Gamma^u}{{\rm d}z {\rm d}x_\ell} &=& 
12 z (1-z) S(\mu_2,m_b) 
\Biggl\{1 + \frac{\alpha_s C_F}{4 \pi} 
\nn
&&
\hspace{-2cm}\times\left[ \ln^2 \frac{\mu_2^2}{m_b^2} - (4 \ln z - 5) \ln \frac{\mu_2^2}{m_b^2} -2 H(z,x_\ell) \right]\Biggr\},
\label{finaldoublediff1}
\end{eqnarray}
with
\begin{eqnarray}
H(z,x_\ell) &=& \ln^2(1-x_\ell) - \left(2\ln z-\frac{7}{2}\right) \ln(1-x_\ell)  + \ln^2 z 
\nn
&&\hspace{-1cm}
+ 2 {\rm Li}_2(1-z) + \frac{(3z-1)\ln z}{2(1-z)} + \frac{2 \pi^2}{3} + \frac{5}{2}.
\end{eqnarray}
As for the $B \to X_s \gamma$ case (section \ref{sec:diffs1}) we have neglected the running below the scale  $\mu_2$ and the dependence on $\mu_2$ in Eq.~(\ref{finaldoublediff1}) cancels. Expanding $S(\mu_2,m_b)$ to first order in $\alpha_s$ we reproduce the result obtained previously obtained by De~Fazio and Neubert \cite{dFN}. 

\subsubsection{$B \to X_u \ell \bar \nu$ for $m_b(1-x_{\ell}) \sim \lqcd$}

As for the decay $B \to X_s \gamma$ the expansion of the bilocal operator can not be performed in this region of phase space and differential decay rates in this region of phase space are given in terms of a convolution of a perturbatively calculable coefficient and the shape function. Using the results of Section \ref{sec:final} for the $W_i^u$ in the shape function region we find
\begin{eqnarray}
\frac{1}{\Gamma_0^u} \frac{{\rm d}\Gamma^u}{{\rm d}z {\rm d}x_\ell} &=& 12 \, z (1-z) S(\mu_2,m_b) \Biggl[ F(p^+,\mu_2)
\\
&&
\hspace{-1cm}
+\frac{\alpha_s C_F}{4 \pi} \int_0^{p^+} \!\!\!\!dk^+ G_u(p^+-k^+,\mu_2) f(k^+,\mu_2) \Biggl], \nonumber
\end{eqnarray}
where $p^+ = m_b(1-x_\ell)$,
\begin{eqnarray}
F(p^+,\mu) = \int_0^{p^+} dk^+ f(k^+,\mu),
\end{eqnarray}
and
\begin{eqnarray}
G_u(\omega,\mu) &=& 2\ln^2 \frac{z \,m_b \,\omega}{\mu^2} - 3 \ln \frac{z \,m_b \,\omega}{\mu^2} - 4 \ln^2 z 
\nn
&&
\hspace{-1cm}- 4 \,{\rm Li}_2(1-z) - 2 \ln \frac{3z-2}{1-z} - 5 - \frac{7\pi^2}{6}.
\end{eqnarray}
Again,  our results disagree with the results presented in Ref.~\cite{dFN}, as we will discuss next.

\subsubsection{Comparison with previous results in the literature}
\label{sec:diffconv}

De~Fazio and Neubert include shape function effects by convoluting the perturbative decay spectrum (i.e. $W_i^{\text{part}}$) with a shape function. This procedure is not correct because there are large perturbative corrections included in the definition of the shape function. Consider, for example, the most singular terms in $W_i^{\text{part}}$, the $\ln p^+/p^+$ terms. These arise from  ${\cal C}^{(1)}(p^+,\mu)  + f^{(1)}_{\text{part}}(p^+,\mu)$. The ${\cal C}^{(1)}$ term gives a coefficient of $4$, and the $ f^{(1)}_{\text{part}}$ term gives a coefficient of $-8$, for a net contribution of $-4$. The order $\alpha_s$ computation of Ref.~\cite{dFN} gives the net result of $-4$, without reference to the scale at which it arises. Our effective theory computation shows that it is generated at two different scales, $4$ is generated from the matching condition at $\mu_2$ onto the shape function operator and $-8$ from the matrix element of the shape function at $\mu_3$.  De~Fazio and Neubert convolute the shape function with the order $\alpha_s$ spectrum, i.e. with the $-4$. Instead, one should convolute it with the order $\alpha_s$ coefficients $C_i^{(f)}$ generated by the time-ordered product of currents at the scale $\mu_2$, which has a coefficient $+4$. The remaining $-8$ is included in the definition of the shape function, since it is produced by taking the matrix element of the shape function operator in the $B$ meson. This somewhat surprising conclusion follows from the fact that the the shape function operator has a non-zero one-loop matrix element between on-shell $b$ quark states, as shown in Eq.~(\ref{4.37a}). In deep inelastic scattering, where the parton distribution operator has zero matrix element between on-shell quark states, the cross-section can be written as the convolution of the order $\alpha_s$ perturbative cross-section with a non-perturbative parton distribution function.

Below the scale $\mu_3$, the entire order $\alpha_s$ coefficient has been removed from
from the operators, and included in the coefficient functions. The theory below $\mu_3$ is an expansion in local operators, which is valid when $p^+ \gg \lqcd$, and so is no longer the shape function region.

\subsection{Renormalization group improved photon energy moments}
\label{sec:rgmom}

Moments of the photon energy spectrum as defined in \cite{KorSter,BFL} are given by (using Eqs.~(\ref{fdef},\ref{20.30},\ref{wgamma}))
\begin{eqnarray}
\lefteqn{M_N^\gamma= \int_0^{m_B/m_b} {\rm d}x_\gamma\, x_\gamma^{N-1}\frac{1}{\Gamma_0} \frac{{\rm d} \Gamma}{{\rm d} x_\gamma} }\nn
&=& \int_0^{m_B/m_b} {\rm d}x_\gamma\, x_\gamma^{N-1} \int_0^{m_b(1-x_\gamma)} {\rm d}r^+\nn
&& C_\gamma\left(m_b(1-x_\gamma)-r^+,\mu\right)  {\me{\bar B}{O_v(r^+)}{\bar B} \over 2m_B}\nn
&=&m_b \int_0^{m_B/m_b} {\rm d}x_\gamma\, x_\gamma^{N-1} \int_{x_\gamma}^{m_B/m_b} {\rm d}y\nn
&& C_\gamma\left(m_b(y-x_\gamma),\mu\right)  {\me{\bar B}{O_v(m_b(1-y))}{\bar B} \over 2m_B} ,\nonumber
\end{eqnarray}
where we have made the change of variables $r^+=m_b(1-y)$.  Note that the physical limit on the photon energy is $m_B/2$, making the upper limit on $x_\gamma$ to be $m_B/m_b$. The limits of integration on $r^+$ arise because the Wilson coefficient vanishes for $r^+ > m_b(1-x_\gamma)$ and $f(r^+)$ vanishes for $r^+ < -\bar \Lambda$. Changing the order of integration, and making the change of variables $x_\gamma =y z$ gives
\begin{eqnarray}
M_N^\gamma &=& \int_0^{m_B/m_b} {\rm d }y\ y^{N-1} {\me{\bar B}{O_v(m_b(1-y))}{\bar B} \over 2m_B} \nn
&& \times \int_0^1 {\rm d}z\ z^{N-1} m_b y C_\gamma\left(m_b y(1-z),\mu\right).\nonumber
\end{eqnarray}
The coefficient $C_\gamma(p^+)$ has terms of the form $\delta (p^+)$, $[1/p^+]_{\dist}$ and $[\ln p^+/p^+]_{\dist}$. Using the results of Appendix~\ref{app:dist}, Eqs.~(\ref{A.07},\ref{A.02}) one can convert to the form
\begin{eqnarray}
C_\gamma(m_b y(1-z),\mu) = C_\gamma(1-z,\mu) + \ln y\  \text{terms},\nonumber
\end{eqnarray}
where $C_\gamma(1-z)$ contains $\delta(1-z)$, $[1/(1-z)]_+$ and $[\ln(1-z)/(1-z)]_+$ terms. In the endpoint region $y \to 1$, $\ln y = \ln \left[1 + (1-y)\right] \to 0$, so the $\ln y$ terms can be dropped. Thus $M_N$ is a product of moments,
\begin{eqnarray}
M_N^\gamma &=& M_N(f) M_N(C_\gamma),
\label{17.07}
\end{eqnarray}
where the moments of the shape function and the $C_\gamma$ are defined by
\begin{eqnarray}
M_N(f) &=& \int_0^{m_B/m_b} {\rm d }y\ y^{N-1} {\me{\bar B}{O_v(m_b(1-y))}{\bar B} \over 2m_B},\nn
M_N(C_\gamma) &=& \int_0^1 {\rm d}z\ z^{N-1} C_\gamma\left(1-z,\mu\right).\nonumber
\end{eqnarray}
The analog of Eq.~(\ref{17.07}) holds for any $W_i^{(f)}$, with $C_\gamma$ replaced by $C_i^{(f)}$,
\begin{eqnarray}
M_N\left(W_i^{(f)} \right) &=& M_N(f)\ M_N\left(C_i^{(f)} \right)
\label{17.08}
\end{eqnarray}
where
\begin{eqnarray}
M_N\left(W_i^{(f)} \right) &=& \int_0^{m_B/m_b} {\rm d}x\ x^{N-1} W(\bn \cdot p,p^+=m_b(1-x)).\nonumber
\end{eqnarray}

The moments of the shape function operator
\begin{eqnarray}
 O_N(\mu) &=&  \int_0^{m_B/m_b} {\rm d}y\ y^{N-1} O_v\left(m_b(1-y) \right), \nonumber
\end{eqnarray}
where
\begin{eqnarray}
 M_N(f) &=&{\me{\bar B}{O_N(\mu)}{\bar B} \over 2m_B},\nonumber
 \end{eqnarray}
satisfy the renormalization group equation for its matrix element
\begin{eqnarray}
&&\hspace{-.4cm}
\mu { {\rm d} \over {\rm d} \mu} \vev{O_N(\mu)} \nn
&=&
{\alpha_s C_F \over \pi} \int_0^{m_B/m_b} {\rm d}y \, y^{N-1}
\int_{-\infty}^\infty  \!\! {\rm d}x \,\,
\vev{O_v\left(m_b (1-x)} \right)
 \nn
&&
\times \bigg\{
\delta\left(x-y \right)  
\nn
&&
\hspace{.2cm}
+2
\left[{\theta\left((x-y) m_b > \xi \right) \over x-y } + \ln {\xi\over \mu}  \delta(x-y) \right] \bigg\}, \nonumber
\end{eqnarray}
using the anomalous dimension Eq.~(\ref{4.31}) with $r^+=m_b(1-y)$ and $\ell^+=m_b(1-x)$. Letting $y=x z$ gives
\begin{eqnarray}
&&\hspace{-.4cm}
\mu { {\rm d} \over {\rm d} \mu} \vev{O_N(\mu)} \nn
&=&
{\alpha_s C_F \over \pi} \int_0^{m_B/m_b} {\rm d}x \, x^{N-1}\vev{O_v\left(m_b (1-x)} \right)
\int_0^1 {\rm d}z z^{N-1} \,\,
 \nn&&
\times \bigg\{
\left(1- 2 \ln \frac{\mu}{m_b x}  \right)\delta\left(1-z \right)  
+2
\left[{2 \over 1-z }\right]_+ \bigg\}.\nonumber
\end{eqnarray}
Since we are looking at large moments, $x \to 1$ and $\ln x \to 0$, so that
\begin{eqnarray}
\mu { {\rm d} \over {\rm d} \mu} \vev{O_N(\mu)}&=& - \gamma_N(\mu)\vev{O_N(\mu)},
\label{17.11}
\end{eqnarray}
where the anomalous dimension $\gamma_N(\mu)$ is, at one loop,
\begin{eqnarray}
\gamma_N &=&
- {\alpha_s C_F \over \pi} \left\{
\left(1- 2 \ln {\mu \over m_b }  \right)  - 2 H_{N-1} \right\},\nonumber
\end{eqnarray}
which for large $N$ is
\begin{eqnarray}
\gamma_N \to - {\alpha_s C_F \over \pi} 
\left(1- 2 \ln {\mu \bN \over m_b }  \right) .
\label{4.38}
\end{eqnarray}
This agrees with the results of \cite{BFL}.  The complete NLO anomalous dimension is given by including the two-loop $\ln N$ part of the anomalous dimension, which is given by the relation Eq.~(\ref{20.35}),
\begin{eqnarray}
\gamma_N &=&  4 B  \frac{\alpha_s^2(\mu)C_F }{(2\pi)^2}
  \,\ln\left( {\bN \mu \over m_b} \right),
\end{eqnarray}
where $B$ is given in Eq.~(\ref{Bdef}).

The solution to this renormalization group equation is given by
\begin{eqnarray}
\vev{O_N(\mu_3)} &=& S_f^{-1}(\mu_3,\mu_2) \vev{O_N(\mu_2)},
\end{eqnarray}
where the $N$ dependent scale factor is
\begin{eqnarray}
S_f(\mu_3,\mu_2) &=&  \exp\Bigg[\frac{r_{f0}(z)}{\alpha_s(\mu_3)} 
  + r_{f1}(z) \bigg]  ,
\label{17.12}
\end{eqnarray}
with
\begin{eqnarray}
r_{f0}(z) &=& -\frac{8\pi C_F}{\beta_0^2}
 \:\Big[ \frac{1}{z} -1 + \ln z \Big] \,,\nn
r_{f1}(z) &=& -\frac{2 C_F\beta_1}{\beta_0^3} \Big[ 1 -z + \ln z 
-\frac12 \ln^2 z
  \Big] 
  \nn&&
+ \frac{2 C_F}{\beta_0} \Big[2 H_{N-1}-2 \ln \bN - 1 \Big] \ln z  
  \nn
  &&
  - \frac{4 C_F B}{\beta_0^2}  \Big[ z -1- \ln z \Big]\,,
\label{fdef3}\end{eqnarray}
where
\begin{eqnarray}
  z &=& {\alpha_s(\mu_2)\over \alpha_s(\mu_3)} . 
 \end{eqnarray}

Combining these results we obtain for the renormalization group improved moments of $W_i^{(f)}$ \begin{eqnarray}
M_N\left[W_i^{(f)} \right] &=& M_N[f(\mu_2)]\ M_N\left[C_i^{(f)}(\mu_2) \right],\nn
M_N\left[C_i^{(f)}(\mu_2) \right] &=& S(\mu_2,\mu_1)  \bigg[a_i^{(f)} M_N\left[{\cal C}(w,\mu_2) \right]\nn
&& + \frac{\alpha_s(m_b) C_F}{4 \pi} b_i^{(f)}   \bigg],\nn
M_N\left[{\cal C}(w,\mu_2) \right]  &=& 1 + {\alpha_s(\mu_2) C_F \over 4 \pi}
M_N\left[{\cal C}^{(1)}(w,\mu_2) \right],\nn
M_N[f(\mu_2)] &=& S_f (\mu_3,\mu_2) M_N[f(\mu_3)] .\nonumber
\end{eqnarray}
At the partonic level,
\begin{eqnarray}
M_N[f(\mu_3)] &=& 1 + { \alpha_s(\mu_3) C_F \over 4 \pi} M_N[f^{(1)}(\mu_3)].\nonumber
\end{eqnarray}
The moments of the perturbative coefficients are, using Eq.~(\ref{A.04}),
\begin{eqnarray}
M_N\left[{\cal C}^{(1)}(w,\mu_2) \right] &=& 4 \sum_{j=1}^{N-1} {H_j \over j} - \left(4 \ln{m_b \bn \cdot p \over \mu_2^2} -3 \right)H_{N-1}\nn
&&\hspace{-1cm}+ \left( 2  \ln^2{m_b \bn \cdot p \over \mu_2^2}-3  \ln{m_b \bn \cdot p \over \mu_2^2} + 7 - \pi^2 \right) , \nn
M_N[f^{(1)}(\mu_3)] &=& -8 \sum_{j=1}^{N-1} {H_j \over j}- 4 \left(\ln{\mu_3^2 \over m_b^2} -1 \right)H_{N-1}\nn
&&+ \left( -  \ln^2{ \mu_3^2 \over m_b^2 } + 2 \ln {\mu_3^2 \over m_b^2} - {\pi^2 \over 6} \right) .\nn
\label{17.21}
\end{eqnarray}

These expressions can be used to obtain the renormalization group improved moments for the $B \to X_s \gamma$ spectrum. Using $a_\gamma^{(s)}=1$,
\begin{eqnarray}
M_N \left[ \frac{1}{\Gamma_0^s}\frac{{\rm d}\Gamma}{{\rm d}x_\gamma}  \right]
&=& M_N\left[ W_\gamma^{(s)}\right]  \nn
&=& S(\mu_2,\mu_1)S_f(\mu_3,\mu_2) M_N[f(\mu_3)] \nn
&& \hspace{-3cm}  \times \Biggl\{ 1+ {C_F \over 4 \pi} \biggl[\alpha_s(m_b) b_\gamma^{(s)}+ \alpha_s(\mu_2)M_N\left[{\cal C}^{(1)}(w,\mu_2) \right] \biggr]\Biggr\}\nn
&&+{ \alpha_s(m_b) C_F \over 4 \pi} M_N\left[ E \right].
\label{momans}
\end{eqnarray}
The computation to leading order in the SCET expansion parameter $\lambda$ does not include the terms $E(x)$ with vanishing moments as $N \to \infty$. We have included these terms so that we can compare the renormalization group improved moments with the fixed order result even for small moments. Since $E(x)$ is generated in the traditional  operator product expansion at the scale $m_b$, we have included these terms with $\alpha_s(m_b)$. We have not included the running of local operators below $\mu_3$. As discussed below, the regime $\mu_3 > \mu_4$ exists only for the first few moments, and for these, the local operators have zero anomalous dimension.

\begin{figure}
\includegraphics[width=9cm]{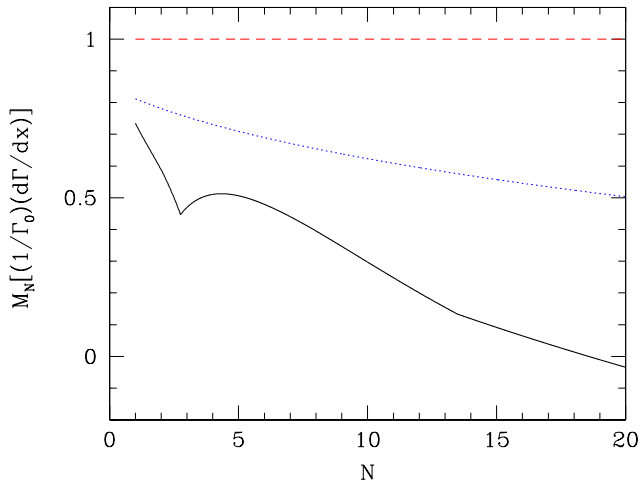}
\caption{The first twenty moments $M_N$ of the $B \to X_s \gamma$ photon spectrum $(1/{\Gamma_0^s})({{\rm d}\Gamma}/{{\rm d}x_\gamma})$ --- renormalization group improved (solid  black), order $\alpha_s$ (dotted blue), and tree level (dashed red). The kink in the blue curve at $N \approx 2.75$ is the point where  $\mu_3=\mu_4$, and the kink at $N \approx 13.5$ is where $\mu_2=\mu_4$.
\label{fig:mom}}
\end{figure}

We can now compare the order $\alpha_s$ moments with the renormalization group improved moments. For simplicity, we will compare the two parton model results, i.e.\ we will use Eq.~(\ref{17.21}) for the moments of the shape function. Non-perturbative effects can be included in $M_N(f)$ if so desired. We choose $\mu_1=m_b$ and $\mu_4=1$~GeV, which are fixed. $\mu_2=m_b/\sqrt{\bN}$ and $\mu_3=m_b/\bN$ vary with the moment. For $N \agt 3$, $\mu_3 \le \mu_4$. In this case, the regime of local operators does not exist. One runs the shape function down to the scale $\mu_4$, and computes its matrix elements in the $B$ meson state at $\mu_4$, so one can effectively set $\mu_3=\mu_4$ in Eq.~(\ref{momans}). For even larger moments, $N \agt 14$, $\mu_2 \le \mu_4$. In this case, the shape function regime does not exist. One runs the SCET currents down to $\mu_4$ and then takes matrix elements of the time ordered product in the $B$ meson state. In this case, one can effectively set $\mu_2=\mu_3=\mu_4$ in Eq.~(\ref{momans}). The order $\alpha_s$ and renormalization group improved moments are show in Fig.~\ref{fig:mom}.
The renormalization group resummation produces a significant correction to the moments, and cannot be neglected.

The large moments in the perturbation theory calculation shown in Fig.~\ref{fig:mom} become negative. This a reflection of the breakdown of perturbation theory. In obtaining Fig.~\ref{fig:mom}, we have stopped the renormalization group evolution at the scale $\mu_4$, but still computed the $B$ matrix element using perturbation theory. The matrix elements in Eq.~(\ref{17.21}) contain $\ln N$ terms, which are an indication of the sensitivity to the lower scales $m_b/\sqrt{\bar N}$ and $m_b/{\bar N}$, and cause a breakdown of perturbation theory for large $N$. The correct computation for large moments involves taking the non-perturbative matrix elements in the $B$ meson state. In the shape function region ($\mu_3 > \mu_4$ but $\mu_2 < \mu_4$), one sets $\mu_3=\mu_4$ in Eq.~(\ref{momans}) and uses the non-perturbative results for the shape function moments $M_N[f(\mu_3)]$. For $\mu_2 < \mu_4$, one sets $\mu_2=\mu_4$, drops the perturbative corrections ${\cal C}^{(1)}$ generated at the scale $\mu_2$, and uses for $M_N[f(\mu_3)]$ the moments of the non-perturbative matrix element of the time-ordered product of currents.

\subsubsection{The decay spectrum vs. its moments}

The effective theory computation is most naturally done in moment space. The intermediate scales $\mu_2$ and $\mu_3$ are given in terms of the moment $N$ as $m_b/\sqrt{\bar N}$ and $m_b/\bar N$. The matching contributions and running depend on $\mu_2$ and $\mu_3$, which change with $N$. Furthermore, for large $N$, the scales $\mu_{2,3}$ can become smaller than $\lqcd$, and the effective theory computation has to be modified, as discussed above.

We have not found a simple way to give a theoretically correct expression for the spectrum in $x$-space which includes the renormalization group evolution, and at the same time properly takes into account effects such as $\mu_{2,3}$ becoming smaller than $\lqcd$. The simple double integral form Eq.~(\ref{20.30}) is not valid because it has a Landau pole singularity as $x \to 1$.

\section{Conclusions}

We have calculated the hadronic tensor for left handed vector and tensor currents, relevant for the decays $B \to X_u \ell \bar \nu$ and $B \to X_s \gamma$ in a region of phase space where the final hadronic system satisfies $p^+ \ll p^-$. From this hadronic tensor any differential decay rate can be obtained. This calculation has been performed using the soft collinear effective theory, by performing the matching in multiple steps and summing all the logarithms arising from the different scales.  We considered two possible regions, where $p^+ \sim \lqcd$ and where $p^+ \gg \lqcd$.

In the former region, the hadronic tensor is given as a convolution of calculable Wilson coefficients and the lightcone distribution function of the $B$ meson. We calculated these Wilson coefficients to order $\alpha_s$ including the leading and subleading Sudakov logarithms. The resulting expressions, when expanded to order $\alpha_s$ (i.e. without renormalization group improvement), agree with the order $\alpha_s$ results previously given by de~Fazio and Neubert \cite{dFN}. However, we do not agree with the assumption in Ref.~\cite{dFN} that the structure function of the $B$ meson does not contain any perturbative effects, so that shape function effects can be included by convoluting the order $\alpha_s$ decay distributions with the shape function. We have seen that very large perturbative corrections are included in the shape function matrix element Eq.~(\ref{4.37b}) at the scale $\mu_3$. We have also shown that the usual assumption that moments of the lightcone distribution function are given by matrix elements of universal local operators (such as the kinetic or Darwin operators) is incorrect. This relation only holds at the classical level, but fails once quantum corrections are taken into account. 

In the region $p^+ \gg \lqcd$ the structure function can be expanded in terms of local operators and the the non-perturbative physics of the hadronic tenor are parameterized by matrix elements of these universal local operators. The results presented here contain the most singular contributions in the endpoint region, which are enhanced by powers of $m_b/p^+$ relative to terms we have dropped. The matching coefficients for these enhanced non-perturbative contributions are again given to order $\alpha_s$. 

\acknowledgments

We would like to thank Michael Luke, Zoltan Ligeti and Mark Wise for discussions. Some of this work was done at the Aspen Center for Physics. This work was supported in part by the Department of Energy under Contracts DE-FG03-97ER40546 and DE-FG03-92ER40701.

\begin{appendix}

\section{The $\dist$ Distribution}\label{app:dist}

It is convenient to use a modified version of the usual $+$-distribution to regulate singular functions of $p^+$ on the infinite interval $p^+ \in [0,\infty]$. The $\dist$-distribution is defined by
\begin{eqnarray}
\left[ {\theta(p^+) \over p^+} \right]_{\dist} &\equiv& \!\!\!\lim_{\xi \to 0} \Biggl[ {\theta(p^+  >\xi) \over p^+ } +  \delta(p^+) \ln {\xi\over \mu}\Biggr] \nn
\left[ {\ln( p^+/\mu) \ \theta(p^+) \over p^+} \right]_{\dist} &\equiv& \!\!\!\lim_{\xi \to 0} \Biggl[{\ln (p^+/\mu) \ \theta(p^+ >\xi) \over p^+}\nn
&& + \frac 1 2 \delta(p^+) \ln^2 {\xi\over \mu}  \Biggr],\nonumber
\end{eqnarray}
and is closely related to the $\ast$-distribution in Ref.~\cite{dFN}. The $\dist$-distribution
satisfies the relations
\begin{eqnarray}
\lefteqn{\int_{-A}^B {\rm d}p^+\ f(p^+) \left[ \theta(p^+) \over p^+ \right]_{\dist} =}\nn 
&& \int_{0}^B {\rm d}p^+\ { f(p^+) - f(0) \over p^+} + f(0) \ln{ B \over \mu}, \nn
\lefteqn{\int_{-A}^B {\rm d}p^+\ f(p^+) \left[ \ln (p^+/\mu) \ \theta(p^+) \over p^+ \right]_{\dist} =}\nn
&&\int_{0}^B {\rm d}p^+\  { f(p^+) - f(0) \over p^+}\ln {p^+ \over \mu}   +  \frac 1 2 f(0)  \ln^2 {B \over \mu} .\nn
\label{4.12a}
\end{eqnarray}
for $A\ge0$ and $B > 0$. The $\dist$-distribution is defined so that integrals such as Eq.~(\ref{4.12a}) involve logarithms of dimensionless quantities.

The $\dist$-distribution satisfies the scaling property
\begin{eqnarray}
\left[ {\theta(\lambda p^+) \over \lambda p^+} \right]_{\dist} &=& {1\over \lambda} \left[ {\theta(\lambda p^+) \over \lambda p^+} \right]_{\dist}+{\ln \lambda \over \lambda}\delta(p^+),\nn
\left[ {\ln( \lambda p^+/\mu) \ \theta(\lambda p^+) \over \lambda  p^+} \right]_{\dist} &=& 
{1\over \lambda}\left[ {\ln(  p^+/\mu) \ \theta( p^+) \over   p^+} \right]_{\dist}\nn
&&\hspace{-2cm} +{\ln \lambda \over \lambda}  \left[ {\theta(\lambda p^+) \over \lambda p^+} \right]_{\dist} +{\ln^2 \lambda \over 2 \lambda} \delta(p^+).
\label{A.07}
\end{eqnarray}

One can also consider the $\dist$-distribution restricted to a finite interval $p^+ \in [0,\Lambda]$. In this case, one can convert to the usual $+$-distributions defined by
\begin{eqnarray}
&&\int_{0}^\Lambda {\rm d}p^+\ f(p^+) \left[ \theta(p^+) \over p^+ \right]_{+} = \int_{0}^\Lambda {\rm d}p^+\ { f(p^+) - f(0) \over p^+}, \nn
\lefteqn{\int_{0}^\Lambda {\rm d}p^+\ f(p^+) \left[ \ln (p^+/\mu) \ \theta(p^+) \over p^+ \right]_{+}=}\nn
&& \int_{0}^B {\rm d}p^+\ { f(p^+) - f(0) \over p^+} \ln {p^+ \over \mu}  .\nonumber
\end{eqnarray}
The conversion is
\begin{eqnarray}
\left[ \theta(p^+) \over p^+ \right]_{\dist} &=& \left[ \theta(p^+) \over p^+ \right]_{+}
+ \delta(p^+) \ln{\Lambda \over \mu}, \nn
\left[ \ln (p^+/\mu) \ \theta(p^+) \over p^+ \right]_{\dist} &=& \left[ \ln (p^+/\mu) \ \theta(p^+) \over p^+ \right]_{+}
+ \frac 1 2 \delta(p^+) \ln^2{\Lambda \over \mu}\nn
&=& \left[ \ln (p^+/\Lambda) \ \theta(p^+) \over p^+ \right]_{+} 
 \nn
&&
+\ln{\Lambda \over \mu}
\left[\ \theta(p^+) \over p^+ \right]_{+}+ \frac 1 2 \delta(p^+) \ln^2{\Lambda \over \mu},\nn
\label{A.02}
\end{eqnarray}
which depends on the size of the interval $\Lambda$. We will need this conversion formula with $\Lambda=m_b$ and $p^+=m_b(1-z)$,
\begin{eqnarray}
 m_b \left[ \theta(p^+) \over p^+ \right]_{\dist} &=& \left[{1 \over1-z }\right]_{+}
-\frac 1 2 \ln{\mu^2 \over m_b^2 } \delta(1-z) , \nn
 m_b \left[ \ln (p^+/\mu) \ \theta(p^+) \over p^+ \right]_{\dist} 
&=& \left[ \ln (1-z)  \over (1-z) \right]_{+} \nn
&&\hspace{-1.5cm}
-\frac 1 2 \ln{\mu^2 \over m_b^2 }
\left[1\over (1-z) \right]_{+} + \frac 1 8 \delta(1-z) \ln^2{\mu^2 \over m_b^2 }.\nonumber
\end{eqnarray}
Using this, Eqs.~(\ref{3.15},\ref{4.37b}) become
\begin{eqnarray}
\mathcal{C}^{(1)}(q^+,\mu_2)  &=& 4 \left[ \ln (1-z)  \over (1-z) \right]_{+} + \left(4 \ln{m_b \bn \cdot p \over \mu_2^2} -3 \right) \left[{1 \over1-z }\right]_{+}\nn
&& \hspace{-1cm} + \left( 2  \ln^2{m_b \bn \cdot p \over \mu_2^2}-3  \ln{m_b \bn \cdot p \over \mu_2^2} + 7 - \pi^2 \right) \delta(1-z),\nn
f_{\text{part}}^{(1)}(k^+,\mu_3) &=& -8 \left[ \ln (1-z)  \over (1-z) \right]_{+} + 4 \left(\ln{\mu_3^2 \over m_b^2} -1 \right) \left[{1 \over1-z }\right]_{+}\nn
&&  + \left( -  \ln^2{ \mu_3^2 \over m_b^2 } + 2 \ln {\mu_3^2 \over m_b^2} - {\pi^2 \over 6} \right) \delta(1-z).\nn
\label{A.04}
\end{eqnarray}

\end{appendix}

\end{document}